\documentclass{article}
\usepackage{epsf}
\usepackage{emulateapj,pstricks}
\input{psfig}

\lefthead{David et al.}
\righthead{ROSAT PSPC OBSERVATIONS OF THE RICHEST ($R \geq 2$) ACO CLUSTERS}

\slugcomment{Accepted by {\em The Astrophysical Journal}}
\def\ref#1{\noindent\hangindent=24.0pt\hangafter=1{#1}\par}
\def\la{\hbox{\rlap{$<$}\lower.5ex\hbox{$\sim$}\ }}
\def\ga{\hbox{\rlap{$>$}\lower.5ex\hbox{$\sim$}\ }}
\def\lax    {{_<\atop^{\sim}}}

\def\ergs   {\rm{~ergs~s^{-1}}}
\def\ergcms   {\rm{~ergs~cm^{-2}~s^{-1}}}
\def\etal   {{\it et~al.}}

\begin{document}

\lefthead{David et al.}
\righthead{ROSAT PSPC OBSERVATIONS OF THE RICHEST ($R \geq 2$) ACO CLUSTERS}

\slugcomment{Accepted by {\em The Astrophysical Journal}}

\title{
ROSAT PSPC OBSERVATIONS OF THE RICHEST ($R \geq 2$) ACO CLUSTERS}

\author{Laurence P. David, William Forman, and Christine Jones}
\affil{Harvard-Smithsonian Center for Astrophysics}
\affil{60 Garden Street}
\affil{Cambridge, MA 02138}

\bigskip

\begin{abstract}

\noindent
We have compiled an X-ray catalog of optically selected
rich clusters of galaxies observed by the 
PSPC during the pointed GO phase of the ROSAT mission. This paper contains
a systematic X-ray analysis of 150 clusters with an optical richness 
classification of $R \geq 2$ from the ACO catalog (Abell, Corwin, and Olowin 1989). 
All clusters were observed within $45^{\prime}$ of the optical
axis of the telescope during pointed PSPC observations.
For each cluster, we calculate:
the net 0.5-2.0~keV PSPC count rate (or $4 \sigma$ upper limit) in
a 1~Mpc radius aperture,
0.5-2.0~keV flux and luminosity, bolometric luminosity, 
and X-ray centroid. The 
cluster sample is then used to examine correlations between the X-ray 
and optical properties of clusters, derive the X-ray luminosity 
function of clusters with different optical classifications, and obtain 
a quantitative estimate of contamination
(i.e, the fraction of clusters with an optical richness 
significantly overestimated due to interloping galaxies) in the ACO 
catalog.

\indent
Due to the large field of view of the PSPC, many rich clusters were
serendipitously observed during the GO phase of the ROSAT mission.
Of the 150 clusters in our sample, 82 were observed serendipitously,
and 68 were targeted observations.  The overall detection rate of 
serendipitously observed clusters is quite high at 76\%. 
However, the detection rate is sensitive to the optical properties of 
clusters, and the details of the optical selection process.
For example, all serendipitously observed Bautz-Morgan Type I and I-II 
clusters are detected, while only 71\% of Bautz-Morgan
Type II, II-III, and III clusters are detected.  Beyond $z \approx 0.1$, 
83\% of the observed Abell clusters are detected, compared
to only 60\% of southern ACO clusters.  Due to the long integration
times in pointed PSPC observations, the typical X-ray luminosity
threshold for detection is quite low at $\sim 10^{43} \ergs$ for 
clusters within z=0.2.  This luminosity is more 
characteristic of an X-ray luminous group rather than a rich cluster.  
The nondetected clusters must therefore be either highly unrelaxed systems
or have an optical richness that is significantly overestimated due to 
contamination by interloping galaxies.  
We show that the later possibility is more likely since low X-ray luminosity 
clusters are preferentially found in denser cluster environments
compared with X-ray luminous clusters.  This contrast in cluster
environment makes low X-ray luminosity clusters more susceptible to
galaxy contamination than luminous systems.
The fraction of clusters less luminous than an X-ray luminous group is 
thus a direct quantitative measure of contamination in the ACO catalog.  
We also find that the X-ray luminosity function of 
Abell clusters is inconsistent with that of southern 
ACO clusters.   Only by comparing $R \geq 2$ Abell
clusters with $R \geq 3$ southern
ACO clusters can we obtain consistency in their X-ray properties.
This indicates that the optical richness of southern ACO clusters
is overestimated by about one richness class.  

\noindent
{\it Subject headings:} galaxies: clusters: general

\end{abstract}

\noindent

\vfill

\section{INTRODUCTION}

X-ray observations of clusters of galaxies over the past
20 years have proven invaluable for determining many fundamental properties 
of clusters. Such observations provide essential diagnostics for 
constraining the history and dynamical evolution of individual
clusters along with models governing the evolution of the universe
as a whole.  The ROSAT observatory (Tr\"{u}mper 1983),
which was launched on June 1, 1990, 
has had a tremendous impact on our knowledge of clusters of galaxies.  
ROSAT conducted the first all-sky survey by an imaging X-ray telescope
in 4 segments between July 1990 and Aug 1991.
This survey produced two major X-ray catalogs of clusters: the ROSAT
Brightest Cluster Sample (BCS; Ebeling $\etal$ 1998) and 
the X-Ray Brightest Abell Cluster Sample (XBACS; Ebeling $\etal$ 1996).
Aside from the 6 months during which the ROSAT All-sky Survey (RASS) was 
conducted,  all other observing time was allocated to guest observers.  
The ROSAT PSPC carried out nearly 6000 observations
during its 4 years of operation in the GO phase of the 
ROSAT mission.  Excluding the shortest observations (integration
times less than 50 seconds) and a series of observations that
were used to fill in the ROSAT All-Sky Survey,
leaves 5003 pointed PSPC observations that covered 16.5\% of the sky.
Of these, 4883 PSPC observations are now available to the public
through, e.g., the HEASARC at GSFC. 
This archive contains a wealth of information
about the X-ray properties of many astronomical objects. 
In this paper, we present the
results of a systematic analysis of pointed PSPC observations of rich 
clusters of galaxies.

Cross-correlating the positions of optically rich
clusters in the ACO catalog (Abell, Corwin, and Olowin 1989),
with the ROSAT Master Observation Catalog shows that 
882 ACO clusters were observed during the 5003 pointed 
PSPC observations that are presently available through the HEASARC.
Most of these observations were serendipitous, and not requested pointings.
The primary advantage of pointed PSPC observations compared with 
ROSAT all-sky survey (RASS) data, is the much longer integration time.  
The typical exposure time for a pointed PSPC observation is 10,000 seconds,
compared with 400 seconds in the RASS.  
Of course, the main advantage of the
RASS over the pointed PSPC phase is the nearly complete
sky coverage.  The longer integration times obtained during 
pointed PSPC observations
have several advantages relevant to the data analysis:
1) lower flux limits for detection, 2) better discrimination between point-like 
and extended sources, and 3) less source confusion.
For these reasons, we have compiled a catalog of 
the 150 optically richest ACO clusters (richness class $R \geq 2$)
observed by the PSPC during the GO phase of the ROSAT mission.

This paper is organized in the following manner.  In section 2, we present
our cluster sample, data reduction techniques, 
and X-ray detection rates.  We also discuss in $\oint 2$
how the X-ray detection rates depend on the optical properties of rich clusters.
In section 3 we compare the X-ray fluxes determined
from pointed PSPC with those obtained from the RASS data for 
all clusters in common with our
sample and the XBACS.  Section 4 investigates correlations
between the X-ray and optical properties of rich clusters. The X-ray 
luminosity function (both parametric and nonparametric) of the entire 
sample and several sub-samples (based on optical properties)
is derived in section 5. Section
6 presents an investigation of contamination 
in the ACO catalog, and finally, the main results are summarized in
section 7.

\section{DATA REDUCTION}

\subsection{2.1 The Sample}

Our sample contains all ACO clusters
with a richness class of $R \geq 2$ ($N_{gal} \geq 80$) that 
lie within $45^{\prime}$ of the optical axis of a pointed PSPC observation.
This selection criterion ensures that the central 1~h$^{-1}_{50}$~Mpc of a cluster 
lies within the full PSPC field of view for all clusters beyond $z = 0.04$.
We use the HEASARC Browse search engine to cross-correlate the cluster
positions in the ACO catalog with the PSPC master observation log.
This search produced 82 serendipitously observed ACO
clusters and 68 targeted clusters (see Table 1).
The mean exposure time for the 150 observed clusters is 9.6 ksec, which is approximately
25 times the mean exposure time in the RASS.

To determine if our sample is representative of the general 
optical properties of ACO clusters,
we compare the distribution of optical richness ($N_{gal}$),  Bautz-Morgan
Type, Rood-Sastry Type, and distance class in Figure 1.
Bautz-Morgan Types are available for 142 of the clusters in the PSPC sample
and we compare their distribution with that of 
Abell's statistical sample (Leir and Van den Berg 1977).
Rood-Sastry classifications and their relative abundances
are available for all clusters in Abell's (1958)

\vskip 1.0in

\begin{center}

TABLE 2

Comparison of Optical Properties (KS test results)

\begin{tabular}{l|ccc}
\hline\hline
Property & Targeted & Serendipitous & Entire Sample \\
\hline
$N_{gal}$ & 0.0026 & 0.97 & 0.064 \\
BM Type & $3 \times 10^{-5}$ & 0.98 & $3 \times 10^{-3}$ \\
RS Type& $3 \times 10^{-3}$ & 0.87 & 0.01 \\
D Class& $4 \times 10^{-6}$ & 0.99 & 0.04 \\
\hline
\end{tabular}

Notes: This table give the probability that the targeted, serendipitous,
and entire PSPC samples were drawn from the same parent population of 
clusters as the
ACO catalog, based on optical richness, Bautz-Morgan
Type, Rood Sastry Type, and Distance class.
\end{center}

\noindent
original catalog from Struble and Rood (1987).  
Figure 1 shows that the targeted observations are biased toward nearby, rich,
Bautz-Morgan Type I, cD clusters, while the optical properties of 
the serendipitously observed clusters reflect the general properties
of ACO clusters.  This impression is confirmed using a 
KS test to determine the statistical significance of these differences
(see Table 2). For example, Bautz-Morgan Type I clusters only comprise
3.4\% of Abell's statistical sample, compared with
26\% of the targeted observations of clusters. Due to the contrasting
optical properties of the targeted
and serendipitous samples, 
these samples will be analyzed separately below.

\subsection{Analysis Techniques}

All PSPC data are screened and reduced using the software package
developed by Snowden $\etal$ (1994) for the analysis of extended objects.
The first 15 seconds of each observation interval is excluded
due to possible aspect uncertainties while the target
is being acquired.  In addition, all time intervals with master veto rates
greater than 170~cts~s$^{-1}$ are excised.  Images, exposure
maps, exposure corrected count rate images, and 
exposure corrected count rate error images
are generated for the resulting good time 
intervals for each of 4 highest energy bands 
(R4-R7) covering the energy range 0.51-2.01~keV (hereafter 0.5-2.0~keV).
The images are blocked to a scale of $15^{\prime \prime}$ per pixel.
For clusters with multiple PSPC exposures 
(i.e., when the ACO cluster centroid is within the 
central $45^{\prime}$ radius region of more than one PSPC 
observation), all observations are aligned and co-added.

Every 0.5-2.0~keV exposure corrected
count rate image is then inspected for the presence of an extended source
within one-third of an Abell radius around the optical position of 
the cluster given in ACO.  If extended emission is detected (based on the 
PSPC PSF of a 5~keV thermal source at the appropriate angle off-axis) than the 
centroid of the X-ray emission determined by the PROS task {\it xexamine}
is used as the center of the source circle.  Since
many of the clusters in this sample are among the X-ray brightest clusters
in the sky, the X-ray identification is usually quite obvious.
In observations without a detected extended source within one-third of an Abell radius,
the center of the source circle is positioned at the ACO optical position.
The total counts are then extracted using a circular source region
with a radius of 1~Mpc in the rest frame of the 
cluster, excluding the emission from all unresolved sources.
The luminosity and angular distances are computed from measured
redshifts, if available, otherwise, redshifts are estimated from the
$m_{10}$ vs. $z$ relation given in Ebeling $\etal$ (1996).
All distant dependent quantities are computed assuming 
$\rm{H_0} = 50$~km$^{-1}$~s$^{-1}$~Mpc$^{-1}$ and $q_0 = 0.5$.
The background rate is determined from several regions in the same 
observation that are free of obscuration by the PSPC window support structure
and beyond 3~Mpc of the cluster centroid.
For the few low redshift clusters that fill the field of view 
(e.g., A426, A1367, A1656), a background rate of 
$3.5 \times 10^{-4}$~cts~s$^{-1}$~arcmin$^{-2}$ is used,
which is the median value in the remainder of the PSPC observations.

The net count rates and errors in the 0.5-2.0~keV
energy band are then determined using the standard propagation of errors
(see below).   
To convert from count rates to fluxes we 
use previously measured temperatures, if available, from
David $\etal$ (1993). Otherwise, we perform an iterative 
procedure to estimate the temperature based on the 
relation between X-ray luminosity and gas temperature 
given in David $\etal$  The 0.5-2.0~keV
and bolometric fluxes are determined assuming a Raymond thermal
plasma model with the measured or estimated gas temperature, 30\% solar abundance,
and the galactic hydrogen column density along the line
of sight toward the cluster from Stark $\etal$ (1992).
The resulting X-ray properties of the cluster sample are given
in Table 1.

\subsection{Sensitivity and Detection Rates}

The flux limit for detecting extended sources depends
on the source morphology, redshift, exposure time, and background rate.
The net count rate of a cluster, $R_s$,  is simply 
$R_s = R_t - R_b (A_s / A_b)$, where
$R_t$ and $A_s$ are the total count rate and area in the source circle, 
and $R_b$ and $A_b$ are 
the background count rate and area in the background region.
The error in $R_s$ is then, 
$\sigma_s^2 = \sigma_t^2 + (A_s / A_B )^2 \sigma_b^2$.
Figure 2 shows the $3 \sigma$ flux limit for a source radius
of 1~Mpc, a background rate of $3.0 \times 10^{-4}$cts~s$^{-1}$~arcmin$^{-2}$,
and a typical background region with $A_b = 300$~arcmin$^2$.  
Flux limits are shown for exposure times of 
400~sec (typical of the RASS), 10~ksec (the mean in our sample),
and 50~ksec (representative of some of the longest PSPC observations). 
Also shown in Figure 2 is the $3 \sigma$ flux limit for a point source with 
no background in the source circle (9 net counts). As the redshift
increases, $A_s / A_b$ decreases, and the flux limit for an 
extended source approaches that of a point source.  Figure 2
shows that typical flux limits in pointed PSPC observations are 
an order of magnitude below that possible in the RASS.
The flux limit for detecting clusters at z=0.3 is approximately 
30 times less than that of clusters at z=0.02. This corresponds to an increase
in luminosity of only a factor of 10 between z=0.02 and 0.3, 
compared with a factor of 220 for point sources at the detection 
threshold.  This shows that extended sources detected using 
a fixed physical source radius are much less affected by Malmquist 
bias than point sources.

The detection efficiency of the clusters in our sample is
summarized in Table 3.
Nearly all of the targeted clusters are detected (64/68).
This is not surprising, since the X-ray flux of many
of the targeted clusters was already known from previous X-ray missions
and the requested exposure times guaranteed a detection.  The overall 

\vskip 1.0in

\begin{center}

TABLE 3

Detection Rates

\begin{tabular}{l|cccc}
\hline\hline
Sample& $z < 0.1$ & $0.1 \leq z < 0.2$ & $z \geq 0.2$ & All Redshifts\\
\hline
Targeted& 33/34 & 21/23 & 10/11 & 64/68\\
& (97\%) & (91\%) & (91\%) & (94\%) \\
& & & \\
Serendipitous& 6/6 & 40/53 & 16/23 & 62/82 \\
& (100\%) & (75\%) & (70\%) & (76\%) \\
&&& \\
Abell& 21/22 & 54/64 & 22/28 & 97/114 \\
&(95\%) & (84\%) & (79\%) & (85\%) \\
&&& \\
Southern ACO& 18/18 & 7/12 & 4/6 & 29/36 \\
&(100\%) & (58\%) & (67\%) & (80\%) \\
\hline
\end{tabular}
\end{center}

\noindent
detection rate of serendipitous clusters is also quite
high at 76\%. There is also a slight difference between
the detection rate of Abell (83\%) and southern ACO (60\%)
clusters beyond $z \approx 0.1$. 
Table 4 shows that all of the nondetected serendipitously observed
clusters have late Bautz Morgan Types.
All serendipitously observed clusters with Bautz Morgan Types
of I or I-II, or $R \geq 2$ are detected.
The mean upper limit on the 0.5-2.0~keV luminosity for 
the nondetected serendipitously observed clusters is $1.4 \times 10^{43} \ergs$, 
which is comparable to the X-ray luminosity of the NGC 5044 group of galaxies 
(David $\etal$ 1994).  Optically,
the NGC 5044 group does not even qualify as a dense Hickson group.
Thus, the nondetected clusters are 
either dynamically unrelaxed multi-component systems, or 
have an optical richness that is 
significantly over estimated due to intervening galaxies.  We discuss 
the effects of contamination in the ACO catalog
in $\oint 6$.

For comparison, Ebeling $\etal$ (1993) cross-correlated the positions 
of sources detected by SASS in the individual RASS scans 
with the optical cluster positions in ACO, and found that 
22\% of $R \geq 1$ clusters were detected above a flux limit of  
$7 \times 10^{-13} \ergcms$ in the $0.1-2.4$~keV bandpass.  
Briel and Henry (1993) analyzed the RASS data 
on a complete sample of Abell clusters
within a 561 square degree region at high galactic latitudes with an
average exposure time of 660 sec.  They detected 46\% 
at a significance level for existence greater than $3 \sigma$, 
and computed X-ray fluxes at better than $3 \sigma$ for 36\%. 
The greater detection rate in pointed observations is simply 
due to the greater exposure times.  However, still longer observations will not
necessarily increase the detection rate substantially, since many of
the clusters not detected in pointed observations may not be physically
bound systems.

\section{COMPARISON WITH XBACS}

Ebeling $\etal$ (1996) compiled a sample of 242 ACO clusters (XBACS) detected
in the RASS with unabsorbed 0.1-2.4~keV fluxes greater than 
$5.0 \times 10^{-12} \ergcms$. There are 42 clusters in common between
our two samples.    To make a direct comparison between the 
analysis of the 

\begin{center}

TABLE 4

Detection Rates of Serendipitous Observed Clusters

\begin{tabular}{l|l|r}
\hline\hline
Bautz-Morgan Type & \multicolumn{2}{c}{\rm R=2~~~~  $R \geq 3$} \\
\hline
I, I-II & 8/8 & 1/1 \\
II, II-III, III & 41/59 & 8/10 \\
\hline
\end{tabular}
\end{center}

\vskip 0.3in

\noindent
pointed PSPC observations and the RASS data,
we have to account for differences in the energy
bandpass and extraction radius.  The uncorrected 0.1-2.4~keV count
rates given in Ebeling $\etal$ are derived from the VTP source detection algorithm,
which essentially gives the net count rates within a radius of $r_{VTP}$.  
The corrected count rates include the source flux beyond $r_{VTP}$ (assuming 
a spherically symmetric King model with $\beta=2/3$ and a core radius 
determined from the data) and a statistical correction for the flux 
from unresolved point sources.  The final unabsorbed 0.1-2.4~keV fluxes are 
then derived from the corrected count rates, the galactic hydrogen column 
densities in Stark $\etal$ (1992), and measured temperatures, if available, 
from David $\etal$ (1993). If the gas temperature is unknown, then the 
temperature is estimated from the X-ray luminosity-temperature relation given 
in White, Jones \& Forman (1997).  The best quantity to use for comparison 
between our two samples is the unabsorbed flux, since it is derived from the 
same values for the hydrogen column density.  For a Raymond thermal plasma model,
the ratio of the unabsorbed flux between 0.5-2.0~keV and 0.1-2.4~keV is 
also very insensitive to gas temperature and only varies from 0.60 for a 2~keV 
cluster to 0.62 for a 10~keV cluster. We therefore use a conversion factor 
of 0.61 for all clusters. To interpolate from the total flux 
in Ebeling $\etal$, to the flux within the central 1~Mpc, 
we assume that all clusters can be represented by a spherically 
symmetric $\beta$ model
with $\beta=2/3$, and core radii between 100 and 250~kpc.
For core radii of 100~kpc and 250 kpc, 91\% and 76\% of the total flux 
is contained within the central 1~Mpc, respectively.

A comparison of the unabsorbed 0.1-2.4 keV fluxes in Ebeling $\etal$ with the 
unabsorbed 0.5-2.0~keV fluxes derived from the pointed PSPC 
observations is shown in Figure 3.  The error bars in Figure 3 only 
include the quoted errors on the count rates.
The dashed lines indicate exact agreement for core radii of 100 and 250~kpc. 
This figure shows that there is excellent agreement between the analysis of 
the RASS data presented in Ebeling $\etal$ with the analysis of the pointed 
PSPC observations presented here.  Only 2 clusters (A2069 and A2151) have
measured fluxes that differ at more $3 \sigma$;
however, both of these are double clusters, and the assumption 
of spherical symmetry used to interpolate between the two flux
measurements is certainly inaccurate.

Using the conversion factors given above, the XBACS flux limit
corresponds to a 0.5-2.0~keV flux limit within the central 
1~Mpc of $2.5 \times 10^{-12} \ergcms$.  There are 
three clusters in our sample (A2034, A3223, and A3301) with fluxes 
greater than this value (see Table 3).
A2034 was removed from the XBACS sample due to possible contamination 
from an AGN. The longer (7.0 ksec) pointed observation of A2034
shows that the dominant X-ray component is extended.  
A3223 has a SASS count below the threshold of 
0.1 cts/s required for inclusion in the XBACS.
Morphologically, this is a complicated cluster with several point sources
embedded within extended emission.  A3301 also has a SASS count rate 
less than 0.1 cts/s.   Overall, there are 45 clusters
above the XBACS flux limit in our sample and 42 of these are included in the XBACS.
This implies that the XBACS is 93\% complete, which is within
the 80\% completeness estimate given in Ebeling $\etal$

\section{CORRELATIONS BETWEEN X-RAY AND OPTICAL PROPERTIES}

A correlation between X-ray luminosity and cluster richness 
has been found from Ariel V, Uhuru, HEAO-1,
Einstein, and Exosat observations (McHardy 1978; 
Jones and Forman 1978; Johnson $\etal$ 1983; 
Edge and Stewart 1991; Burg $\etal$ 1994).
There have also been claims for and against a correlation
between $L_x$ and Bautz-Morgan Type (Bahcall 1977; McHardy 1978;
Jones and Forman 1978; Johnson $\etal$ 1983).  
Most of these early samples
contained too few clusters to isolate independent correlations 
between richness and Bautz-Morgan Type, and only contained
the most X-ray luminous clusters.

A scatter plot of $L_x$ versus cluster richness, $N_{gal}$, is shown in Figure 4.
Different symbols are used to indicate early Bautz-Morgan Type
clusters (I and I-II) and late Bautz-Morgan Type clusters 
(II, II-III, and III).
For comparison, we also show in Figure
4, the 0.5-2.0~keV  luminosity of the NGC 5044 group.
While there is a significant amount of scatter in this plot, the 
generalized Kendall's $\tau$ test (which uses the information 
contained in the nondetections) gives a probability of 
only $3 \times 10^{-4}$ that $L_x$ and $N_{gal}$ are uncorrelated for 
the entire sample, and a probability of $< 10^{-4}$ for the serendipitously
observed sample.
Figure 4 shows that the X-ray
luminosity of R=2 clusters varies by a factor of 500,
and completely spans the range from small groups,
to the most X-ray luminous clusters.

The significance of the correlation between 
$L_x$ and $N_{gal}$ actually depends on the Bautz-Morgan Type.
Examination of Figure 4 shows that the X-ray luminosity of
early Bautz-Morgan Type clusters appears to be independent of cluster richness.
This impression is confirmed using the generalized Kendall's $\tau$ test,
which gives an 83\% probability that $L_x$ and $N_{gal}$ are uncorrelated for
the early Bautz-Morgan Type clusters. 
This indicates that the well know correlation between X-ray luminosity and 
richness primarily pertains to late Bautz-Morgan Type clusters.

Based on the entire sample of $R \geq 2$ clusters,
the X-ray luminosity is also strongly correlated with the Bautz-Morgan
Type. There is only a $2 \times 10^{-4}$ probability that  $L_x$ and 
Bautz-Morgan Type are uncorrelated for the entire sample, and a probability
of $2 \times 10^{-3}$ for the serendipitously observed sample.
To differentiate between the dependence of $L_x$ on 
cluster richness and Bautz-Morgan Type, we show scatter
plots of the X-ray luminosities and Bautz-Morgan Types
for just the $R=2$ clusters in Figure 5.
While there is significant scatter in $L_x$ for a given Bautz-Morgan Type, there 
is a systematic decline in $L_x$
between Type I and Type III clusters. The median X-ray luminosity of 
the Type III clusters is almost an order of magnitude less than
that of early Bautz-Morgan
Type clusters.
Using the nonparametric logrank test (which uses the information
contained in the upper-limits), the probability that 
the early Bautz-Morgan Type clusters are culled from the same 
parent population as the 
late Bautz-Morgan Type clusters is less that $10^{-4}$.

These results show that the X-ray luminosity of clusters is
just as strongly correlated with Bautz-Morgan Type as 
with cluster richness.  For a given cluster richness, the X-ray 
luminosity of early Bautz Morgan Type clusters is greater than late 
Bautz-Morgan Type clusters.  This may indicate that cD clusters
are denser and more evolved systems, or that the cluster richness
in late Bautz-Morgan Type clusters is overestimated due
to a greater contamination of interloping galaxies
(see $\oint 6$).

\section{LUMINOSITY FUNCTION}

\subsection{Non-Parametric}

We use the Kaplan-Meier estimator (see, e.g., Feigelson \& Nelson 1985)
to derive the non-parametric representation of the luminosity function.
Clusters at low galactic latitude ($\mid b \mid < 20^{\circ}$)
and beyond $z=0.25$ are excluded from this analysis.
The resulting 0.5-2.0~keV luminosity functions for the 
serendipitous and targeted samples 
are shown in Figure 6a.  It is obvious from this figure
that the targeted observations are biased toward X-ray luminous
clusters. Based on a logrank test, the probability that 
the targeted and serendipitous
samples were drawn from the same parent population is 
only $3 \times 10^{-4}$.  
There is also a significant difference between the X-ray
luminosity function of Abell and southern ACO clusters (see Figure 6b).
A logrank test gives a probability 
of only $9 \times 10^{-3}$  that the northern and southern samples were
drawn from the same parent population.  This probability increases
slightly to $0.015$ if we restrict the comparison to serendipitously observed
clusters. If we only include $R \geq 3$ clusters from the southern 
ACO catalog, than the probability increases significantly to 23\%.
These tests show that the richness of southern ACO clusters 
is overestimated with respect to northern Abell clusters
by approximately 1 richness class.  This will be discussed further 
in $\oint 6$.

In the previous section we, presented scatter plots
of X-ray luminosity vs. cluster richness and
Bautz-Morgan Type. While there are significant 
correlations between these quantities, there is a great
deal of scatter in the plots.  The differences between 
the X-ray luminosities of clusters as a function of 
richness and Bautz-Morgan Type are more impressive
after compiling luminosity functions.
Figure 6c compares the luminosity function of R=2 clusters with $R \geq 3$ clusters.
A logrank test gives a probability of 
$7 \times 10^{-3}$ that these clusters were
culled from the same population.
Figure 6d compares the luminosity function of early Bautz-Morgan Type clusters
with late Bautz-Morgan Type clusters.
A logrank test gives a probability of 
$9 \times 10^{-4}$ that these clusters were
culled from the same population, which is slightly more 
significant than the difference between richness class.

\subsection{Normalization of the Luminosity Function}

To properly normalize the luminosity functions, we need
to determine the volume 
surveyed as a function of X-ray luminosity for each sub-sample. 
For a completely random X-ray sampling of ACO clusters, the proper 
normalization of the luminosity function is the volume 
surveyed in the ACO catalog, or equivalently, the number density 
of ACO clusters.  Scaramella $\etal$ (1991) 
calculated the co-moving number density of $R \geq 1$
Abell and southern ACO clusters to be $8.7 \times 10^{-6}$~h$^3$~Mpc$^{-3}$,
and $12.5 \times 10^{-6}$~h$^3$~Mpc$^{-3}$, respectively.
Weighting these number densities by the solid angles surveyed 
in the Abell (4.47 Sr) and southern ACO catalog (2.77 Sr), 
and the fraction of $R \geq 2$ clusters relative to 
$R \geq 1$ clusters in the two catalogs, gives an average
co-moving number density of 
$3.6 \times 10^{-7}$~h$^3_{50}$~Mpc$^{-3}$ for $R \geq 2$ clusters.

The dependence of the volume surveyed on the X-ray luminosity
of clusters in the targeted sample can be determined through a comparison 
with XBACS (Ebeling $\etal$ 1996).
The XBACS contains all ACO clusters 
detected in the RASS with 0.1-2.4~keV fluxes greater than 
$5.0 \times 10^{-12} \ergcms$.
For this flux limit, XBACS should be volume limited within 
$z \leq 0.25$ for clusters more luminous than $1.5 \times 10^{45} \ergs$
in the 0.1-2.4~keV bandpass.
There are 9 such clusters in the XBACS with $R \geq 2$, and 
8 of these are contained in our targeted sample.
This indicates that our targeted sample is nearly
volume limited for $R \geq 2$ clusters, within $z=0.25$, and
more luminous than $\sim 10^{45} \ergs$.
A lower limit to the luminosity function 
at the high luminosity end of the targeted sample can thus be
obtained by using the volume within $z=0.25$ and 
within the solid angle surveyed in the ACO catalog.
Figure 7 compares the luminosity
function of the volume normalized targeted sample (correcting
for a factor of 9/8 for incompleteness) with the
luminosity function of the ACO number density 
normalized serendipitous sample (correcting for the fraction
of ACO clusters included in the targeted sample).  
This figure shows that the targeted sample
is roughly 5\% volume complete at low luminosities. 
As a consistency check, the completeness of the targeted sample
at low luminosities should just be the fraction (8\%) of all $R \geq 2$
ACO clusters for which pointed PSPC observations were carried
out successfully. Figure 7 shows that the survey volume 
increases with luminosity in the targeted sample, and 
decreases with luminosity in the serendipitous
sample. Thus, the true X-ray luminosity function of $R \geq 2$ 
clusters should connect the low luminosity end of the ACO number 
density normalized serendipitous sample with
the high luminosity end of the volume normalized targeted sample.

The main goal of this section is to derive the X-ray luminosity function
of an optically selected sample of clusters. We have already shown
how the optical properties of clusters in the PSPC sample differ
those in the 
from ACO catalog. However, we can still construct a luminosity function
from the PSPC observations which reflects the luminosity
function of a randomly selected sample of ACO clusters.
To accomplish this, we implement a slight modification to the
Kaplan-Meier estimator, which is given by:

$$\Phi(L_x) = 1 - \prod_{i,L_{x,i} > L_x}^{n} (1 - d_i / n_i)$$

\noindent
where $d_i$ equals the number of detections at $L_{x,i}$
and $n_i$ is the number of clusters with 
$L_x \leq L_{x,i}$. To correct for biases in the optical
properties of the PSPC sample, we write $n_i$ as:

$$n_i = \sum_{j \leq i}  f_{j}^{-1}$$

\noindent
were $f_{j}$ is the under-sampling of a cluster with 
an X-ray luminosity of $L_{x,j}$ based on its 
richness and Bautz-Morgan Type.  For example, there are
35 clusters in the ACO catalog with $R=2$ and Bautz Morgan Type I,
compared to 12 such clusters in the PSPC sample. Thus, each such
cluster in the PSPC sample is treated as 2.9 clusters in the construction of 
the X-ray luminosity function.
Defined in this way, we can construct an X-ray luminosity function
of ACO clusters with the proper abundance of each richness
class and Bautz-Morgan Type.  
The resulting 0.5-2.0~keV and bolometric luminosity functions are shown in 
Figure 8.  By using the above
modification to the Kaplan-Meier estimator, we have successfully
generated a luminosity function that agrees with the ACO 
number density normalized
serendipitous sample at low luminosities, and the
volume normalized targeted sample at high luminosities.

\subsection{Parametric Luminosity Function}

Cluster X-ray luminosity functions derived from HEAO-1 (Kowalski, Ulmer,
and Cruddace 1983), Einstein (Burg $\etal$ 1994), and ROSAT
(Ebeling $\etal$ 1997) observations are best fit by a 
Schechter luminosity function, given by:

$$\Phi (L)dL = N^* \left( {{L} \over {L^*}} \right)^{-\alpha}  exp \left( {{-L} \over {L^*}} \right) d \left( {{L} \over {L^*} } \right) $$

\noindent
We fit our sample using the detections and bounds technique described in Avni
\& Tananbaum (1986).  The normalization is adjusted in the fitting procedure 
to satisfy the observed number density of 
$R \geq 2$ clusters ($N^* = n(R \geq 2) / \Gamma (1-\alpha))$.
The maximum likelihood method gives $\alpha = 0.29 (0.25-0.33)$ and 
$L^* = 3.2 (2.7-3.7) \times 10^{44} \ergs$ for the 0.5-2.0~keV
luminosity function and $\alpha = 0.47 (0.43-0.51)$ and 
$L^* = 2.0 (1.6-2.4) \times 10^{45} \ergs$ for the bolometric luminosity
function. Errors are given at the 90\% confidence limit for one interesting 
parameter ($\chi^2_{min} + 2.71$).  

\vskip 1.5in

\begin{table*}

\begin{center}

TABLE 5 

Comparison of X-Ray Luminosity Functions

\begin{tabular}{ccccccc}
\hline\hline
Sample & $L_{min}$ & 75\% & $L_{median}$ & 25\% &  $L^*$ & $\alpha$  \\
& ($\ergs$) & ($\ergs$) & ($\ergs$) & ($\ergs$) & ($\ergs$) & \\
\hline\hline
PSPC & $4.5 \times 10^{42}$ & $3.3 \times 10^{43}$ & $7.6 \times 10^{43}$ & $2.1 \times 10^{44}$ & $(3.2 \pm 0.5)  \times 10^{44}$ & $0.29 \pm 0.04$  \\
IPC &  $1.5 \times 10^{43}$ & $8.4 \times 10^{43}$ & $1.9 \times 10^{44}$ & $3.9 \times 10^{44}$ & $(4.3 \pm 0.6) \times 10^{44}$ & 0.4 (fixed)  \\
BCS & $3.3 \times 10^{42}$  & $6.9 \times 10^{43}$  & $1.8 \times 10^{44}$  & $4.5 \times 10^{44}$ & $(5.7  \pm 1.1) \times 10^{44}$ & $1.85 \pm 0.09$ \\
BH93 & $3.0 \times 10^{43}$  & $3.5 \times 10^{43}$  & $5.5 \times 10^{43}$  &$1.5 \times 10^{44}$  & - & - \\
\hline
\end{tabular}

\end{center}

Notes: The IPC and BH93 luminosities have been converted to the 0.5-2.0~keV bandpass
for comparison with the PSPC and BCS samples.
Errors are given at the 90\% confidence level for the PSPC sample, and at
the 68\% confidence level for the IPC and the BCS samples. 
The percentiles give the fraction of clusters more luminous than a 
given X-ray luminosity.
\end{table*}

\subsection{Comparison With Previous Luminosity Functions}

The earliest cluster X-ray luminosity functions were derived from Uhuru,
Ariel V, and HEAO-1 data (Schwartz 1978; McHardy 1978; 
Kowalski, Ulmer, and Cruddace 1983).
The main differences between these early cluster samples and the pointed PSPC 
sample are the improved sensitivity and low energy coverage
of the PSPC, which increases the efficiency of detecting low luminosity, 
cool clusters.  
An examination of the the Ariel V and Uhuru catalogs 
(compiled by Jones and Forman 1978)
and the HEAO-1 catalog (Kowalski $\etal$ 1984) shows that the least luminous 
$R \geq 2$ cluster detected in these early X-ray missions was A1367.
The X-ray luminosity of A1367 is only slightly below the mean in our 
larger sample (48th percentile).  While A1367 was 
considered an unusually low luminosity rich cluster
based on the early X-ray missions, our sample shows that the X-ray luminosity of 
A1367 is quite representative of $R \geq 2$ clusters.  

Burg $\etal$ (1994) compiled a sample of 212 Abell clusters observed by the
Einstein IPC and calculated separate luminosity functions for $R=0$, $R=1$, and
$R \geq 2$ clusters.  The X-ray luminosities in Burg $\etal$ are derived from 
the net 0.5-4.5~keV emission within 1~Mpc in the rest frame of the clusters.
Assuming a source spectrum of $kT= 5$~keV and a galactic hydrogen column
density of $N_H = 3 \times 10^{20}$~cm$^{-2}$ gives a conversion factor
of 1.8 between the IPC and PSPC luminosities.
A summary of the
main X-ray properties of the PSPC and IPC samples is given in Table 5.
While the high luminosity end of the luminosity functions are in good
agreement (above the 25th percentile), the PSPC sample has a much greater 
fraction of low luminosity clusters.  There are 16 clusters in the PSPC sample 
less luminous than the least luminous cluster in the IPC sample,
and the median luminosity in the PSPC sample is a factor of 2.5 less 
than that in the IPC sample (after correcting for bandpass effects) . 
Since the IPC sample only contains 28 $R \geq 2$ clusters,
Burg $\etal$ were unable to constrain $\alpha$ and $L^*$ independently and 
choose to fix $\alpha=0.4$ in their fitting procedure. This value was
chosen based on the X-ray luminosity function of optically less rich clusters.
Our best fit
value of $\alpha$ is slightly less, but within the 90\% errors
of the less rich clusters in Burg $\etal$  
The IPC sample consists mostly
of targeted observations which produces the 
agreement with the PSPC sample at high luminosities.
The disagreement between the IPC and PSPC luminosity
functions at low luminosities arises from the large 
fraction of serendipitously observed clusters in the PSPC sample.

The ROSAT Brightest Cluster Sample (BCS) consists of all clusters 
in the northern hemisphere ($\delta \geq 0^{\circ}, \mid b \mid \geq 20^{\circ}$) 
that are detected in the ROSAT all-sky survey 
above a flux limit of $4.4 \times 10^{-12} \ergcms$ (Ebeling $\etal$ 1998).  
The BCS includes ACO clusters, Zwicky clusters, and X-ray bright
extended sources identified by the SASS.  In addition, the BCS includes
all clusters detected serendipitously with the VTP
algorithm within a $2^{\circ}$ square region around all X-ray bright ACO or
Zwicky clusters, as well as around all X-ray bright extended sources detected
by the SASS in the study region. In a separate paper, Ebeling $\etal$ (1997) 
fit a Schechter function to all clusters in the BCS within $z = 0.3$.
The results of their fitting procedure 
are summarized in Table 5 for the 0.5-2.0~keV
bandpass.  At luminosities above $3 \times 10^{44} \ergs$ there
is good agreement between the BCS and PSPC samples, and 
the two estimates of $L^*$ are consistent (after adjusting $L^*$ in the BCS 
to give the flux within 1~Mpc).
This probably results from the correlation
between cluster richness and X-ray luminosity, which ensures that
an optically selected sample of the richest ACO clusters will necessarily
contain a large fraction of the most X-ray luminous clusters.  At low 
luminosities, however,
the luminosity functions of the BCS and PSPC samples are very different.
The luminosity function of the BCS continues to rise at low luminosities
while the PSPC sample essentially turns over, due to the lack of low
X-ray luminosity $R \leq 2$ clusters in the PSPC sample. 
There is also good agreement at low luminosities between the luminosity 
function of the BCS and a sample of optically selected
poor clusters (Burns $\etal$ 1996).
The steep luminosity functions of the BCS and the poor cluster samples
cannot extend to arbitrarily low luminosities since the 
integrated luminosity of 
a Schechter function diverges if $\alpha > 1$.  This requires that 
the composite luminosity function of clusters, groups, and galaxies must turn over at 
lower luminosities.

Finally, we compare our luminosity function with a complete sample of 145 ACO
clusters observed at high galactic latitudes during the RASS (Briel and Henry 1993).
To make a direct comparison with our sample, we extracted the 47 clusters
from the Briel and Henry paper
with $R \geq 2$, and used the Kaplan-Meier estimator to
derive a non-parametric representation of the luminosity function. The 
results are summarized in Table 5.  Briel and Henry give X-ray luminosities
in the 0.5-2.5~keV bandpass, which requires a conversion factor of 
approximately 0.80.  Examination of Table 4 shows that there
is virtually no difference between these two estimates of the X-ray luminosity
function of $R \geq 2$ clusters, even through the selection effects are quite
different.  This comparison shows that our method of correcting for bias
in the pointed PSPC sample produces a luminosity function in good agreement
with a purely random sampling.  Due to the greater sensitivity of 
pointed PSPC observations compared to the RASS, our cluster sample spans a 
factor of 350 in X-ray luminosity compared to a factor of 20 in the 
Briel and Henry sample.

\section{CONTAMINATION IN THE ABELL AND ACO CATALOGS}

The identification of clusters from optical surveys is 
complicated by the presence of foreground and background
galaxies and clusters. There have been many estimates of 
the fraction of clusters in the Abell and ACO catalogs that are 
significantly affected by contamination
(Lucey 1983; Frenk $\etal$ 1990; Struble and Rood 1991;
van Haarlem, Frenk, \& White 1997) and
these estimates vary from 3-40\%.
Most of these contamination estimates are based on
$R \geq 1$ clusters.  Lucey estimated that contamination
in $R \geq 2$ clusters is
less significant by a factor of approximately 3.
Since the emissivity of the hot gas in clusters
scales as $n_e^2$, X-ray observations are much less affected by 
contamination.  The large number of ACO clusters observed during 
pointed PSPC observations permits an accurate assessment of 
contamination in these catalogs.
The main results of sections 2 and 5 relevant to this issue are: 
1) 97\% of targeted Abell and southern ACO clusters and
100\% of serendipitously observed clusters with $z < 0.1$ are detected, 
2) at larger redshifts, 83\% of Abell clusters, 75\% of
serendipitously observed clusters, and 60\% of southern ACO clusters
are detected, 3) all Bautz-Morgan Type I and I-II 
clusters are detected (including 9 serendipitously observed clusters),
4) the X-ray luminosity function of $R \geq 2$ Abell clusters
is significantly different than that of $R \geq 2$ southern ACO clusters,
5) the luminosity function of $R \geq 3$ southern ACO clusters
is comparable to that of $R \geq 2$ Abell clusters.

These results indicate that the presence of a cD galaxy
is a very reliable indication of a dense galactic and gaseous environment.  The 
different X-ray luminosity functions for the Abell and southern
ACO catalogs indicate that the optical richness of southern clusters 
is overestimated by about one richness class.  This result is in 
agreement with Scaramella $\etal$ (1991) who 
found "a tendency for rich clusters ($R \geq 2$) to be classified
richer in ACO than in Abell," by comparing the galaxy counts
of clusters in the overlap region between the two studies.
This discrepancy is probably due to the different 
types of plates and the background subtraction techniques used
in the two studies.  Abell visually searched 
the 103a-E red plates of the Palomar Sky Survey to produce his original
1958 catalog, while UK 1.2m Schmidt telescope IIIa-J plates, 
which are blue-yellow sensitive, were used to compile the ACO catalog.
Abell also derived the galaxy background from the same plate 
in a region apparently "free of clusters," while a global
galaxy background was used in the generation of the ACO catalog. Since many 
clusters are located within superclusters and large 
scale filaments, the use of a global background can artificially increase the
estimated richness of a cluster.

There are 34 clusters in our sample with $L(0.5-2.0~keV) < 3.0 \times 10^{43} \ergs$.
Of these, 22 are in Abell's statistical sample
and 12 are in the southern ACO catalog.
These clusters are good candidates for significant contamination since
this luminosity is comparable to the luminosity of a dense group and a factor
of 50 less than the most X-ray luminous clusters.
While the exact degree of contamination is difficult to estimate,  a reduction
of approximately 2 richness classes may be appropriate for these clusters.
Of the 22 clusters in the
Abell statistical catalog, 2 (A1186 and A2198) are members of 
superclusters identified by
West (1989), 2 (A1515 and A1607) are members of supercluster candidates in
Batuski and Burns (1985), and A1566 is listed as a candidate for contamination
in Struble and Rood (1991). Of the remaining 17 Abell clusters, only A536 has a 
measured redshift
low enough ($z \lax 0.1$) to be included in present supercluster
catalogs.  Five of the low luminosity clusters (A968, A998, A1005, A1046, A1049) 
are contained within two nearby PSPC pointings (40$^{\prime}$ separation). 
Since the measured redshifts of these clusters 
are all between 0.190-0.203, they are probably members of a supercluster.
Of the 12 low luminosity clusters in the southern ACO catalog, 2 (A3559 and A3566)
are members of the Shapley supercluster and one (A3818) is listed as a supercluster
candidate in Batuski $\etal$ (1995).  Of the remaining 9, only A3093 has a measured
redshift that is less than 0.1. There is thus good evidence that a large fraction 
of the low luminosity clusters are members of superclusters and may 
be subject to significant contamination.

We are primarily concerned about the level of possible contamination 
in the low X-ray luminosity sample, and not whether these systems are part of a 
physically bound supercluster.  For the former,
we only need to search for nearby clusters with similar values of $m_{10}$, for the 
later, redshift surveys are required.  We have
calculated the angular separation of the nearest neighboring ACO cluster 
for all clusters in the PSPC sample.  Only clusters with similar values
of $m_{10}$ ($\Delta m_{10} \leq 0.2$) are considered as possible sources of 
contamination.  In figure 7, we plot the cumulative distribution
of angular separations for low and high X-ray luminosity
clusters.  It is obvious from this figure that low X-ray luminosity 
clusters have an excess of nearby rich clusters compared with their 
more X-ray luminous counterparts.  Seventy percent of low luminosity
clusters have a neighboring cluster within $1^{\circ}$, compared to
only 35\% of more X-ray luminous clusters.  A KS test only gives a 
$3 \times 10^{-3}$ probability
that these two distributions were drawn from the same 
population.  It is worth noting that at a redshift of $z \sim 0.1$,
an Abell radius ($1.5 h^{-1}$) corresponds to about 20 arcminutes.  Thus,
the presence of a neighboring cluster within $1^{\circ}$ certainly increases
the possibility of contamination.
This analysis shows that the optical richness of 
the low X-ray luminosity clusters is probably overestimated.
Also, the large observed scatter between the X-ray luminosity and optical richness of
clusters (see Figure 4) is probably due to varying degrees of 
contamination and not due to
intrinsic variations between the X-ray properties of comparably
rich clusters.

We can give a rough quantitative estimate of the fraction of clusters
in the Abell and southern ACO catalogs that suffer from significant contamination
(approximately 2 richness classes)
as the fraction of serendipitously observed clusters with 
$L(0.5-2.0~keV) < 3.0 \times 10^{43} \ergs$.  This gives 
20\% and 35\% for the Abell and southern ACO clusters. The larger fraction
for the ACO clusters reflects the systematic off-set of 1 richness
class previously mentioned.

\section{SUMMARY}

Over the next decade a host of new X-ray telescopes
will be launched into orbit (e.g., AXAF, XMM, Spectrum X-Gamma, 
Astro-E, ABRIXAS).  However, with the exception of ABRIXAS, none of 
these telescopes will carry out an all sky survey or have a 
detector with as large a field of view as the PSPC.  Thus, for the near future, 
the ROSAT PSPC data archive will remain the prime database for searching 
fairly deep X-ray exposures of a substantial fraction of the sky.  
In this paper, we present a systematic data analysis
of the 150 optically richest clusters 
observed by the PSPC during the GO phase of the ROSAT mission
(i.e., richness class 2 or greater in the ACO catalog).  
Due to the long exposure times in pointed PSPC observations,
the detection threshold in most observations is comparable to 
an X-ray luminous group 
($L(0.5-2.0~\rm{keV}) \sim 3 \times 10^{43}$~ergs~s$^{-1}$).
Undetected clusters must therefore be either highly unrelaxed or physically
unbound systems (i.e., have an optical richness that is significantly
overestimated due to contamination by intervening galaxies).
The depth of this sample provides tight constraints for 
comparing the X-ray and optical 
properties of rich clusters, deriving luminosity functions,
and determining the significance of galaxy contamination in the ACO catalog.

The detection probability of an ACO cluster is strongly dependent 
on its optical properties. All serendipitously observed
Bautz Morgan Type I and I-II clusters are detected in PSPC 
observations, while only 71\% of
later Bautz Morgan Type clusters are detected. This result
shows that the presence of a central dominant galaxy 
is a very reliable diagnostic
for dense galactic environments.  The PSPC cluster sample 
shows a strong correlation between X-ray luminosity and
optical richness.  For a given optical richness,
there is also a correlation between X-ray luminosity and Bautz Morgan Type,
with Bautz Morgan Type I clusters being the most luminous.
We also find that the X-ray luminosity of Bautz-Morgan Type I 
clusters is independent of cluster richness, at least for 
$R \geq 2$ clusters.

There are also some discrepancies between the X-ray properties
of clusters in the Abell and southern ACO catalogs.
The X-ray luminosity function of $R \geq 2$ 
clusters in Abell's catalog is 
inconsistent with that of $R \geq 2$ clusters in the southern ACO 
catalog. However, the X-ray luminosity function of 
$R \geq 3$ clusters in the southern catalog
is consistent with $R \geq 2$ clusters in the northern 
catalog.  This indicates that there is a systematic off-set of 
approximately 1 richness class between the northern and southern catalogs,
in agreement with the results of Scaramella $\etal$ (1991).
The level of contamination by intervening galaxies in the Abell and ACO
catalogs can be estimated as the fraction of clusters 
with X-ray luminosities less
than $3.0 \times 10^{43} \ergs$ (typical of an X-ray luminous
group).  Using this criterion, we find that 20\% of Abell clusters
and 35\% of southern ACO clusters suffer from significant 
contamination (at least 2 richness classes).  
The use of a global rather than a local galaxy background in ACO 
can explain the higher richness classification and
greater level of contamination compared with the Abell catalog.
Essentially all of the low redshift ($z<0.1$), 
low X-ray luminosity clusters reside within 
superclusters.  We also show that 70\% of low luminosity
clusters ($L(0.5-2.0~\rm{keV} < 3.0 \times 10^{43} \ergs$)
have a neighboring ACO cluster with comparable galaxy 
magnitudes ($\Delta m_{10} \leq 0.2$) within an angular separation 
of $1^{\circ}$, while only 35\% of more luminous 
clusters have a similar ACO cluster within $1^{\circ}$.

The next generation of X-ray telescopes will acquire a great deal
of detailed information on clusters of galaxies (e.g., gas temperatures,
abundance of heavy elements, and small scale substructure) which 
will complement the PSPC observations and further 
our basic understanding of the formation and evolution of rich clusters.

We are grateful to the referee, H. Ebeling, for a very thorough reading
of an early version of this paper and providing us with the percentiles
on the X-ray luminosity function of the BCS given in Table 5. 
This research was supported by NASA Grants NAG5-2745 and NAS8-39073.


\begin{figure*}[tb]
\psfig{file=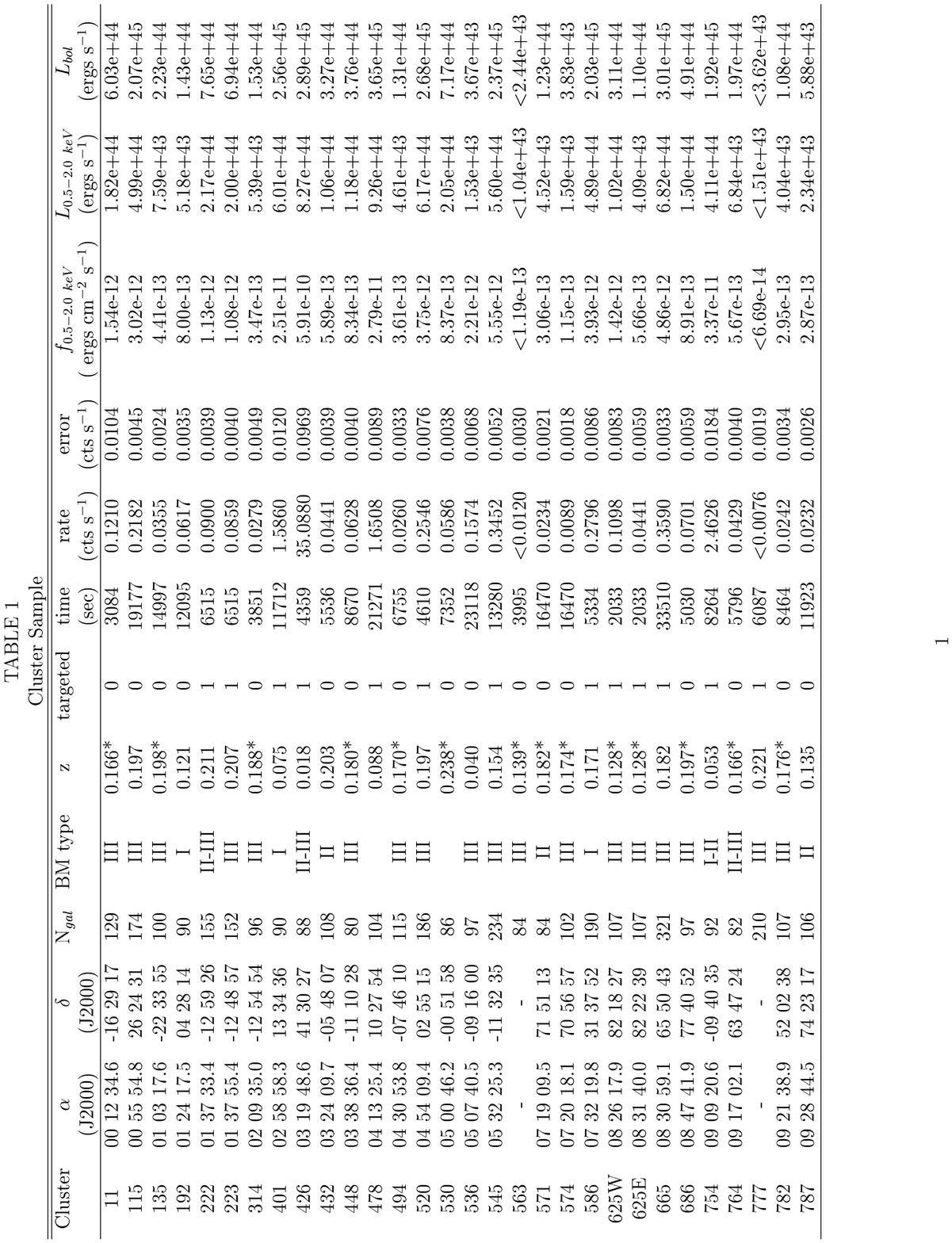,width=7.5in,height=10.0in,angle=180}
\end{figure*}

\begin{figure*}[tb]
\psfig{file=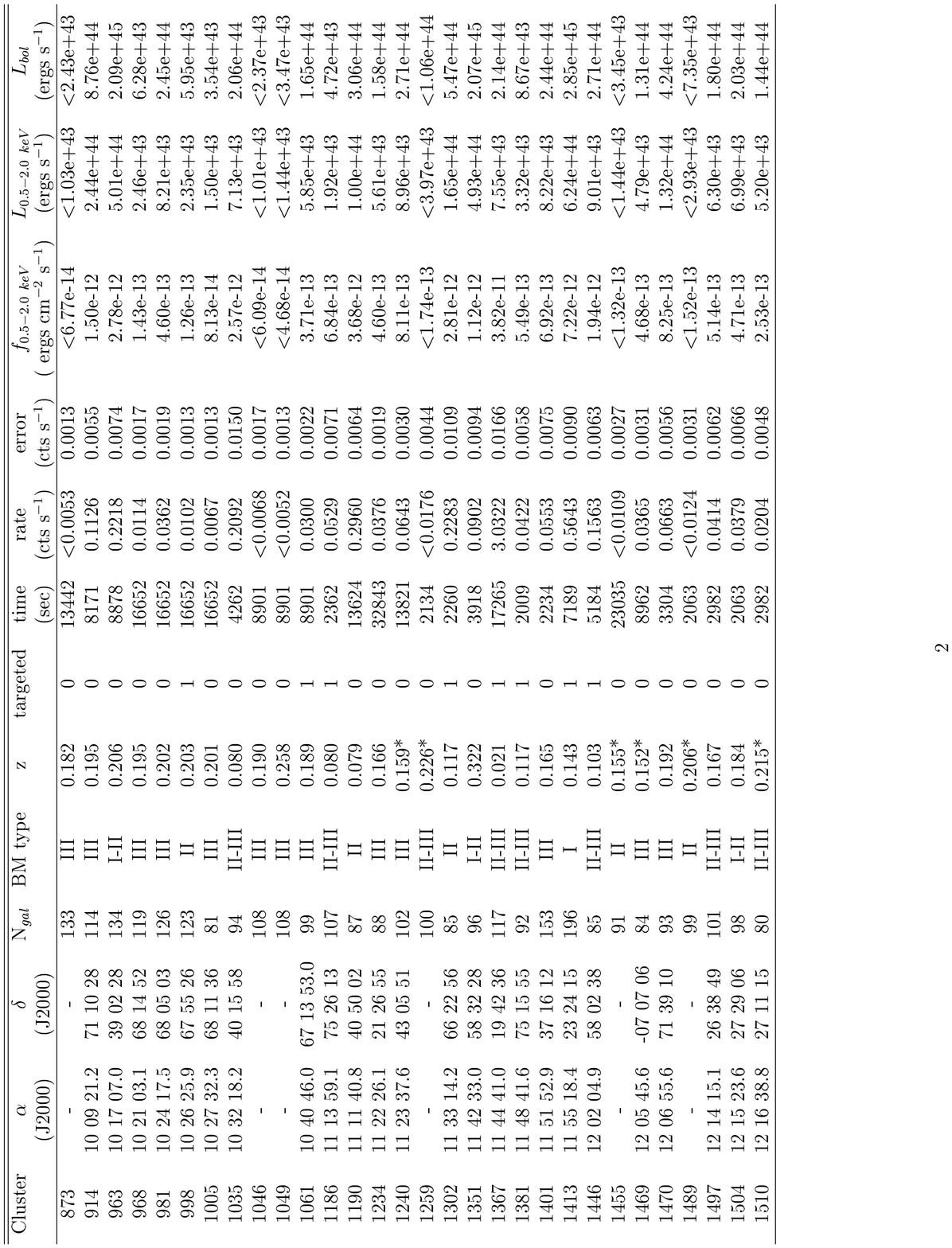,width=7.5in,height=10.0in,angle=180}
\end{figure*}

\begin{figure*}[tb]
\psfig{file=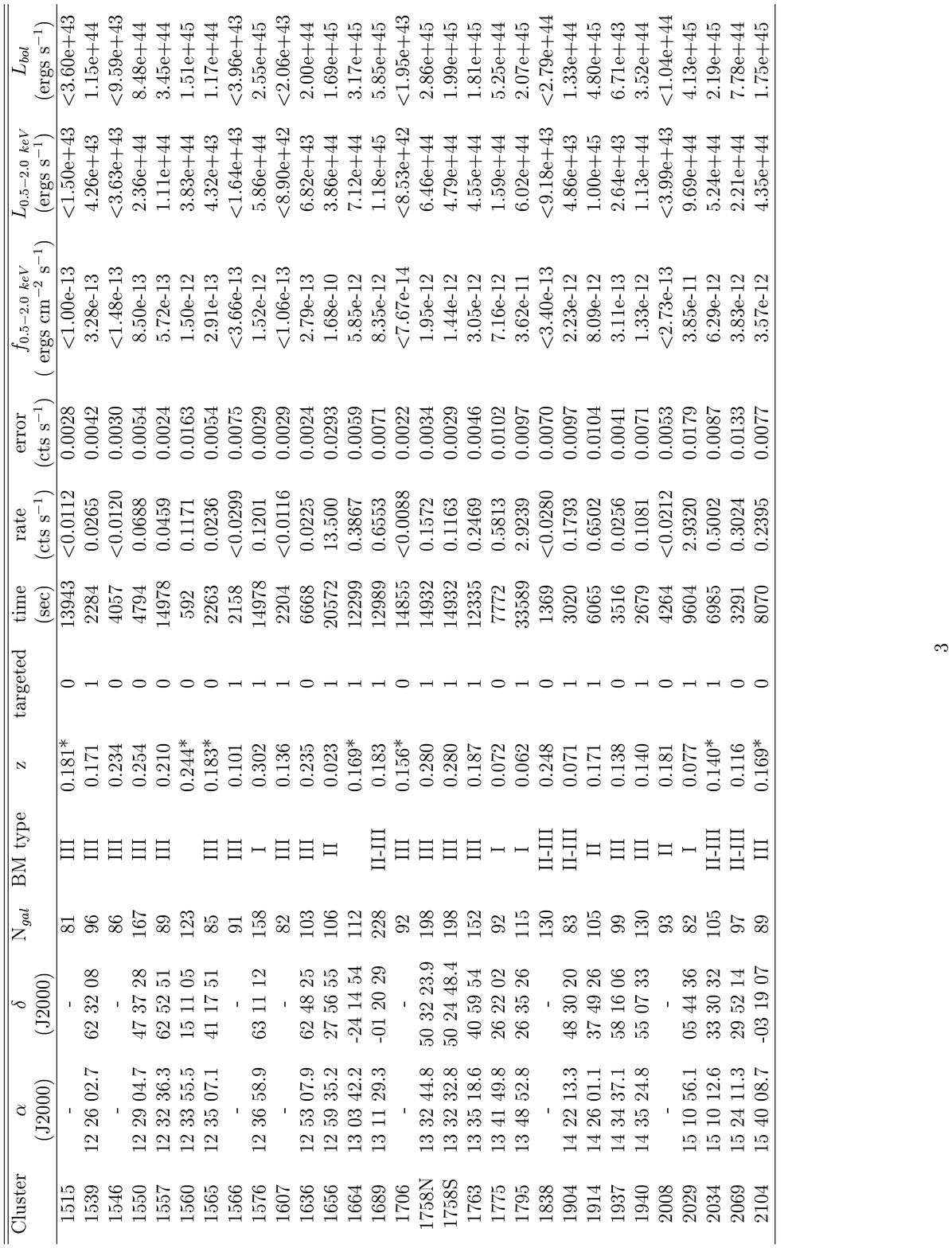,width=7.5in,height=10.0in,angle=180}
\end{figure*}

\begin{figure*}[tb]
\psfig{file=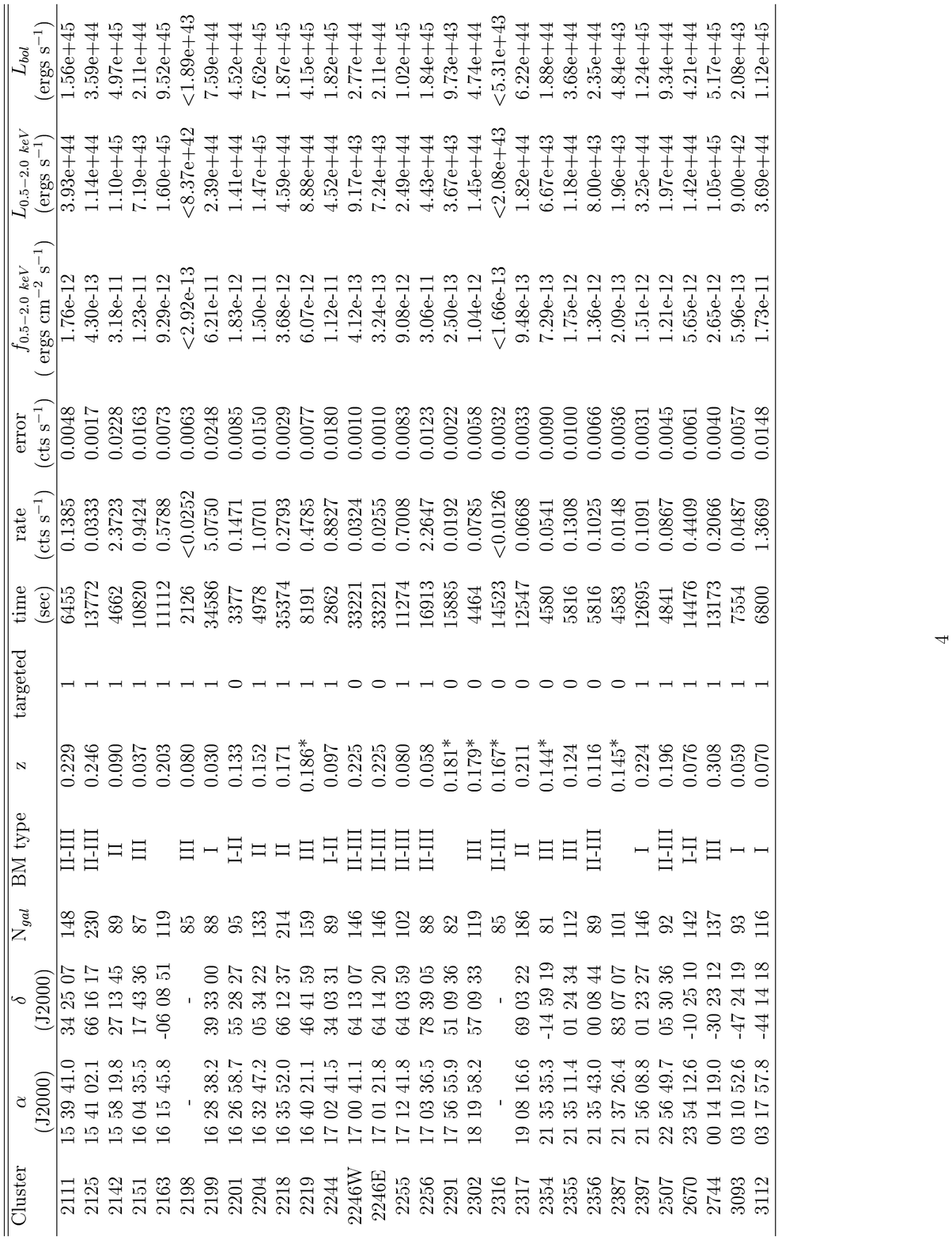,width=7.5in,height=10.0in,angle=180}
\end{figure*}

\begin{figure*}[tb]
\psfig{file=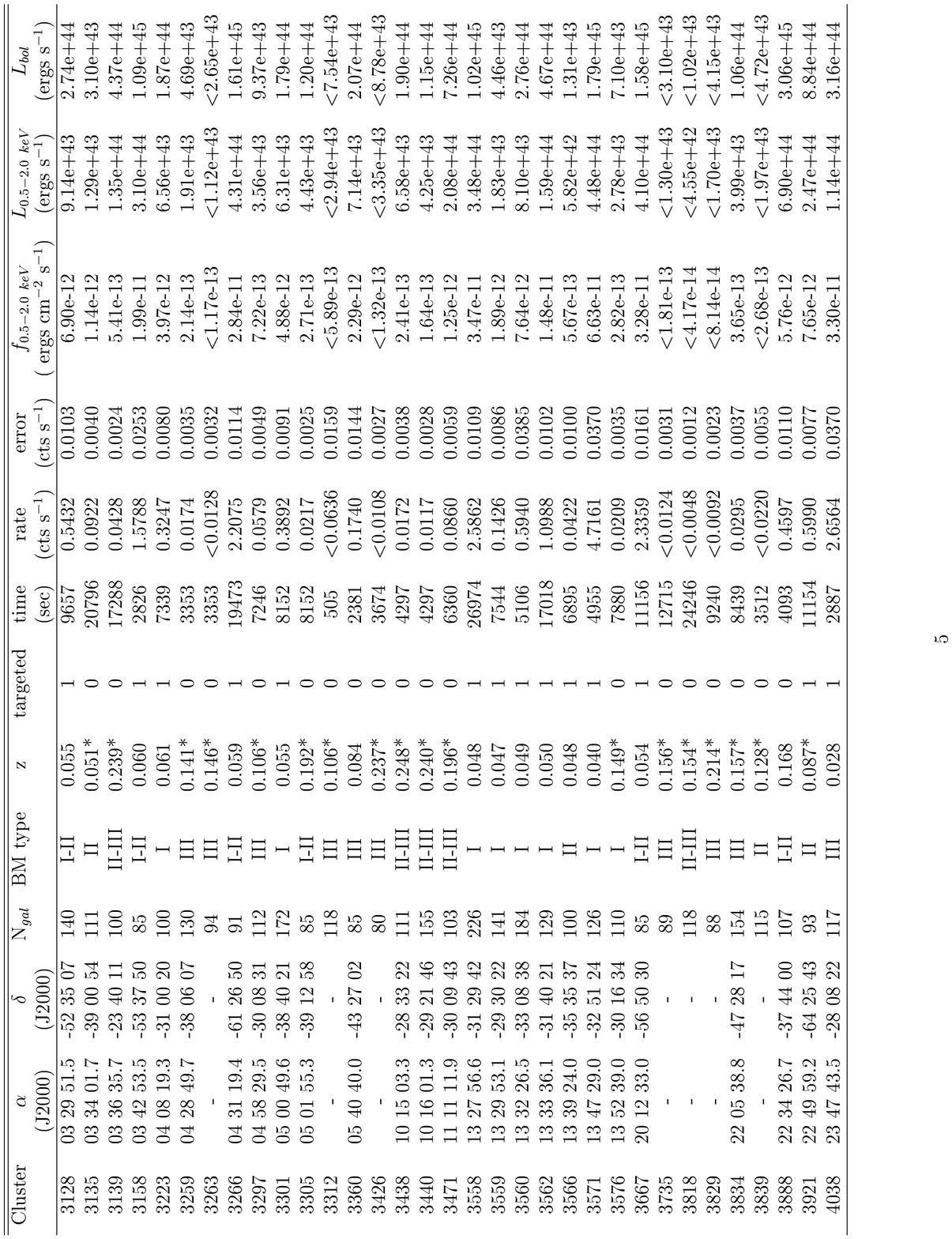,width=7.5in,height=10.0in,angle=180}
\end{figure*}

\begin{figure*}[tb]
\psfig{file=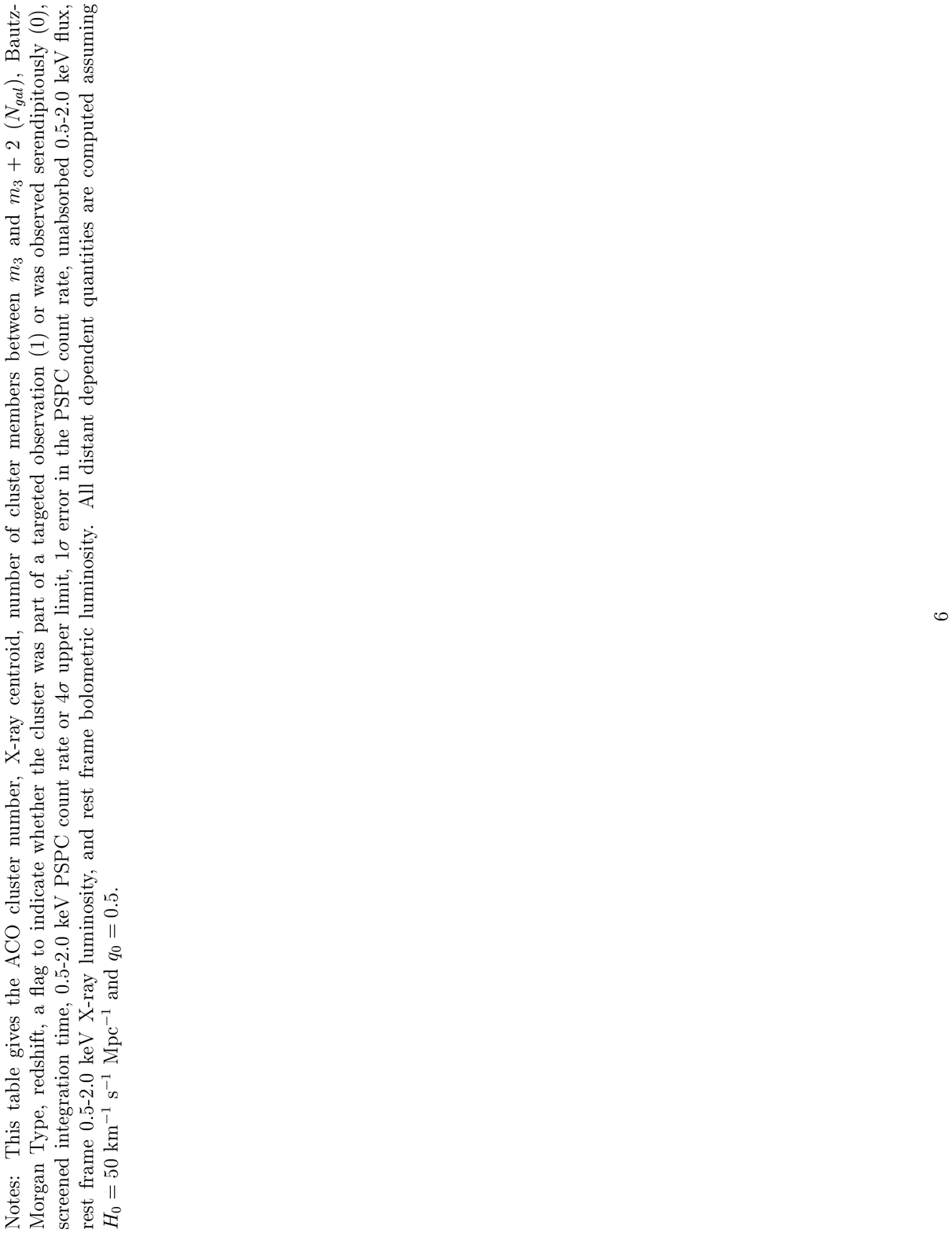,width=7.5in,height=10.0in,angle=180}
\end{figure*}

\begin{figure*}[tb]
\pspicture(0,9.8)(18.5,22.9)

\rput[tl]{0}(0.5,27.2){\epsfxsize=8.5cm
\epsffile{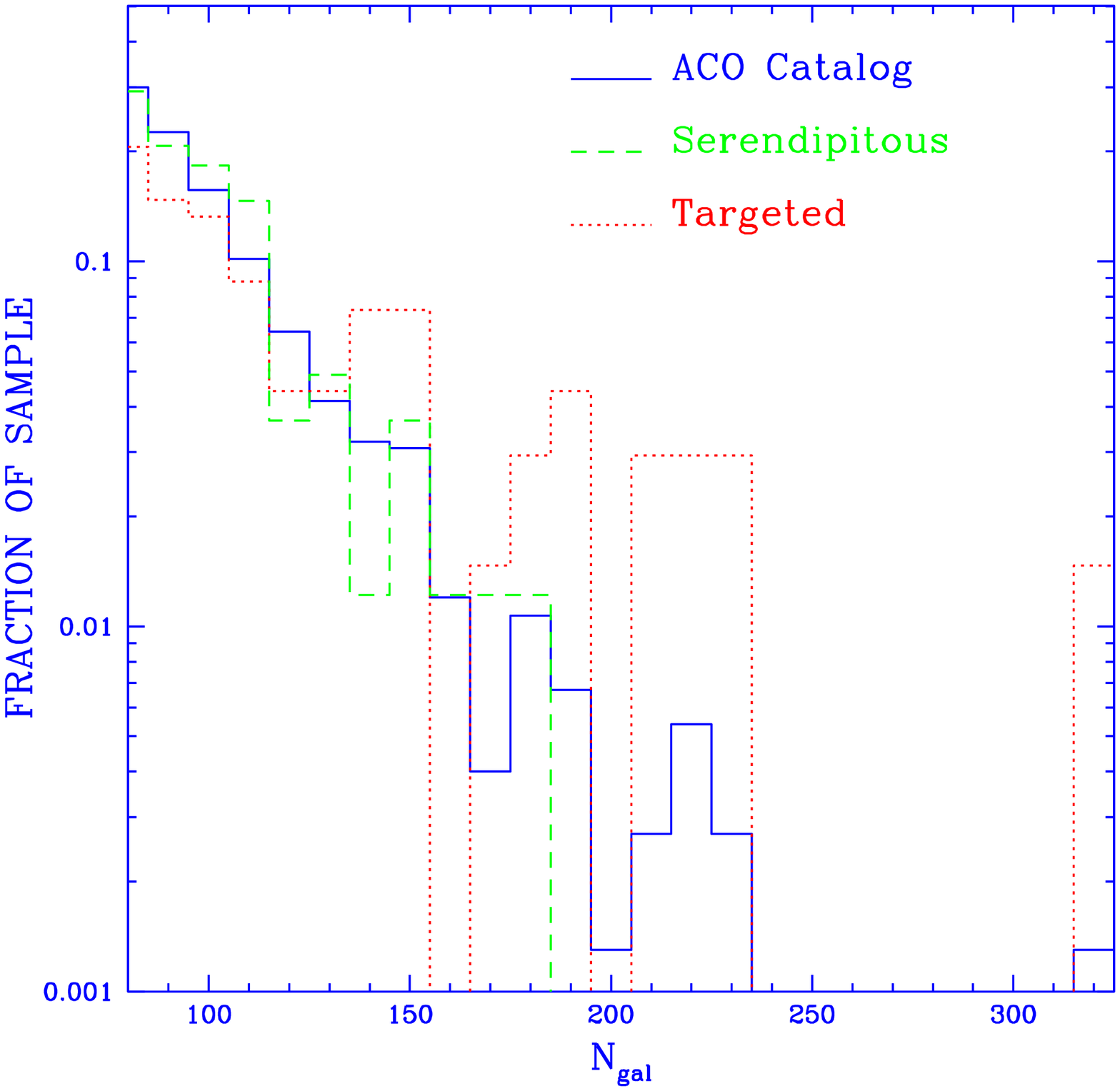}}

\rput[tl]{0}(0.75,16.2){\epsfxsize=7.75cm
\epsffile[30 470 530 678]{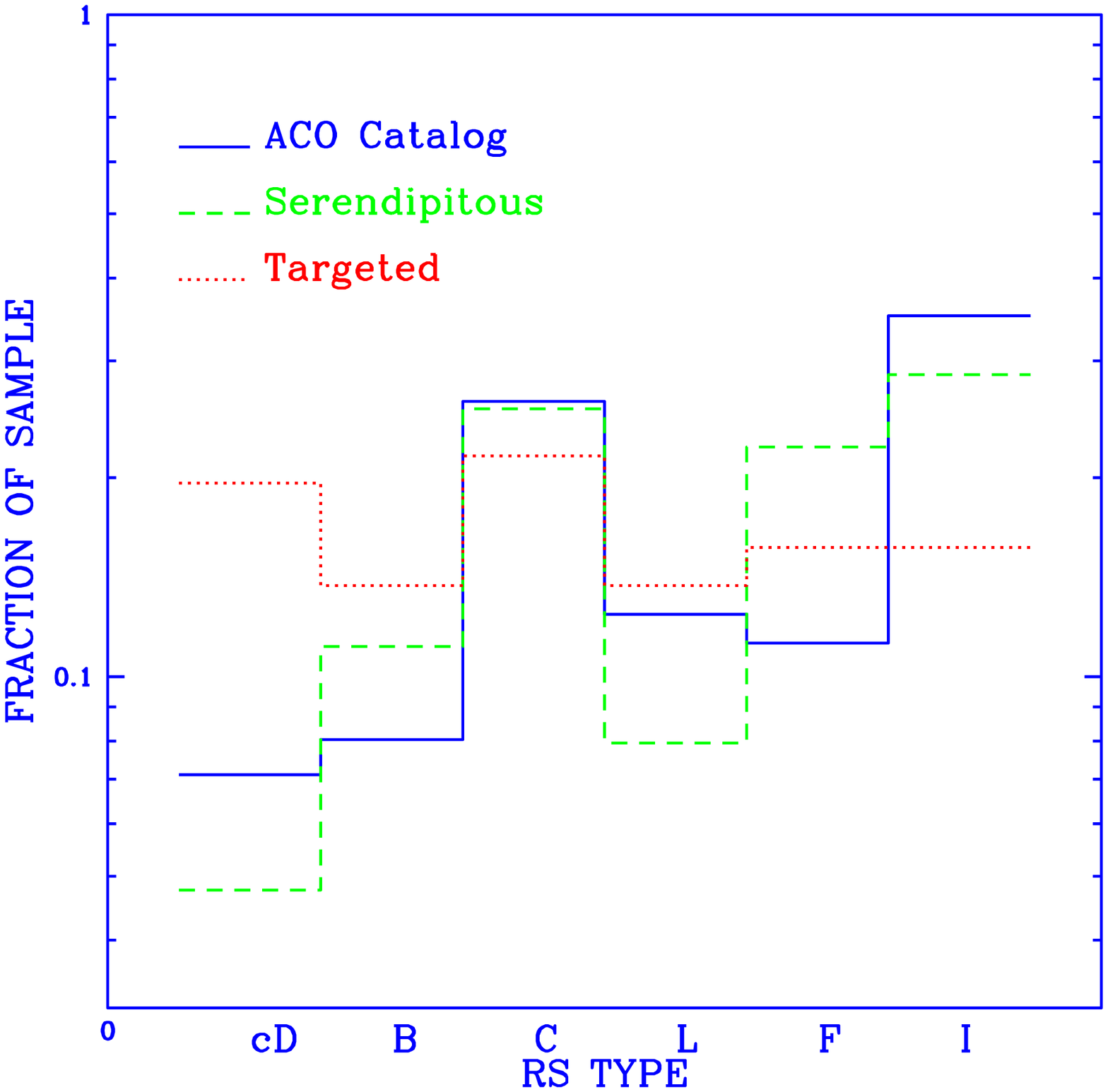}}

\rput[tl]{0}(9.5,27.2){\epsfxsize=8.5cm
\epsffile{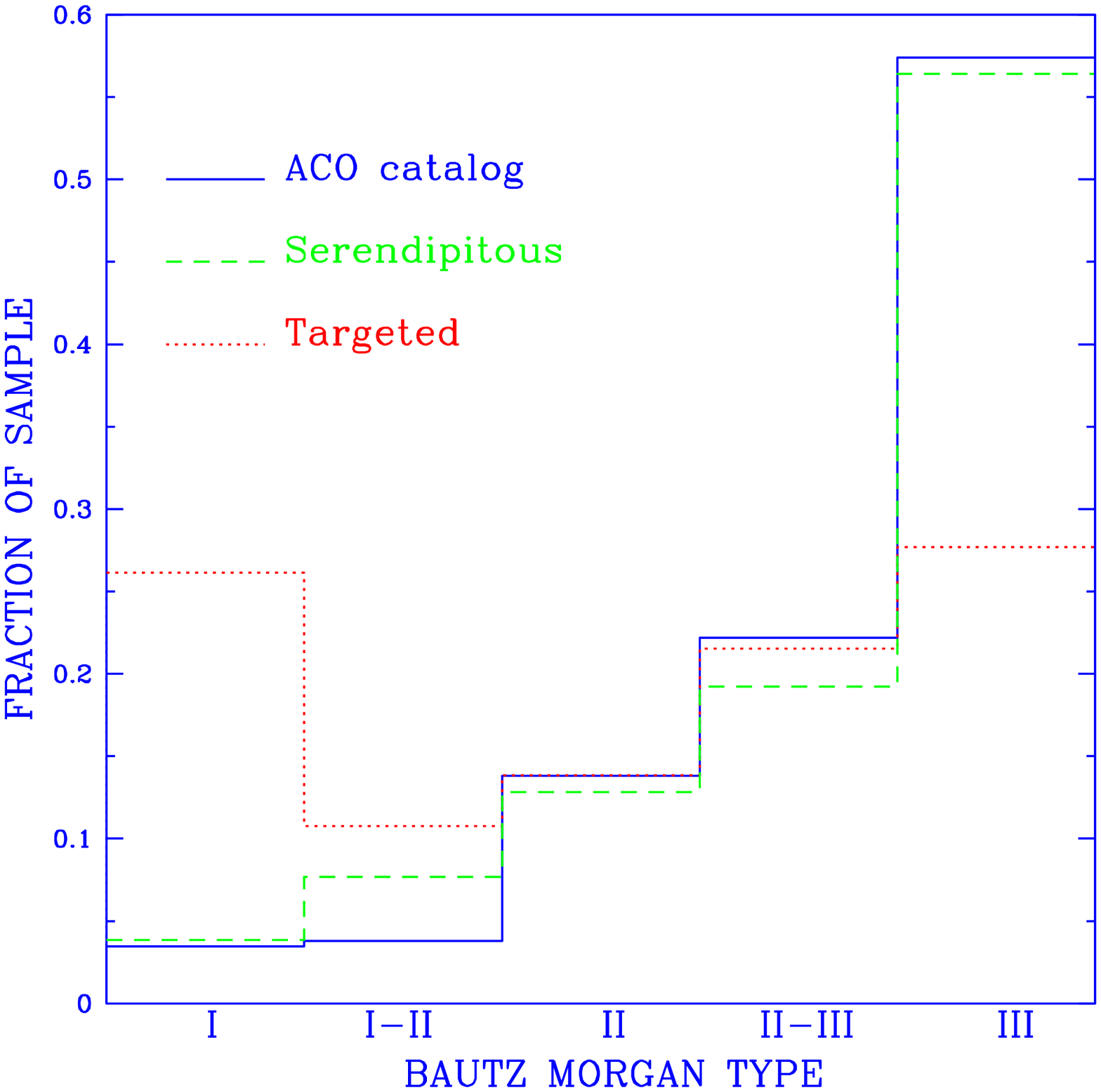}}

\rput[tl]{0}(9.75,16.2){\epsfxsize=7.75cm
\epsffile[30 470 530 678]{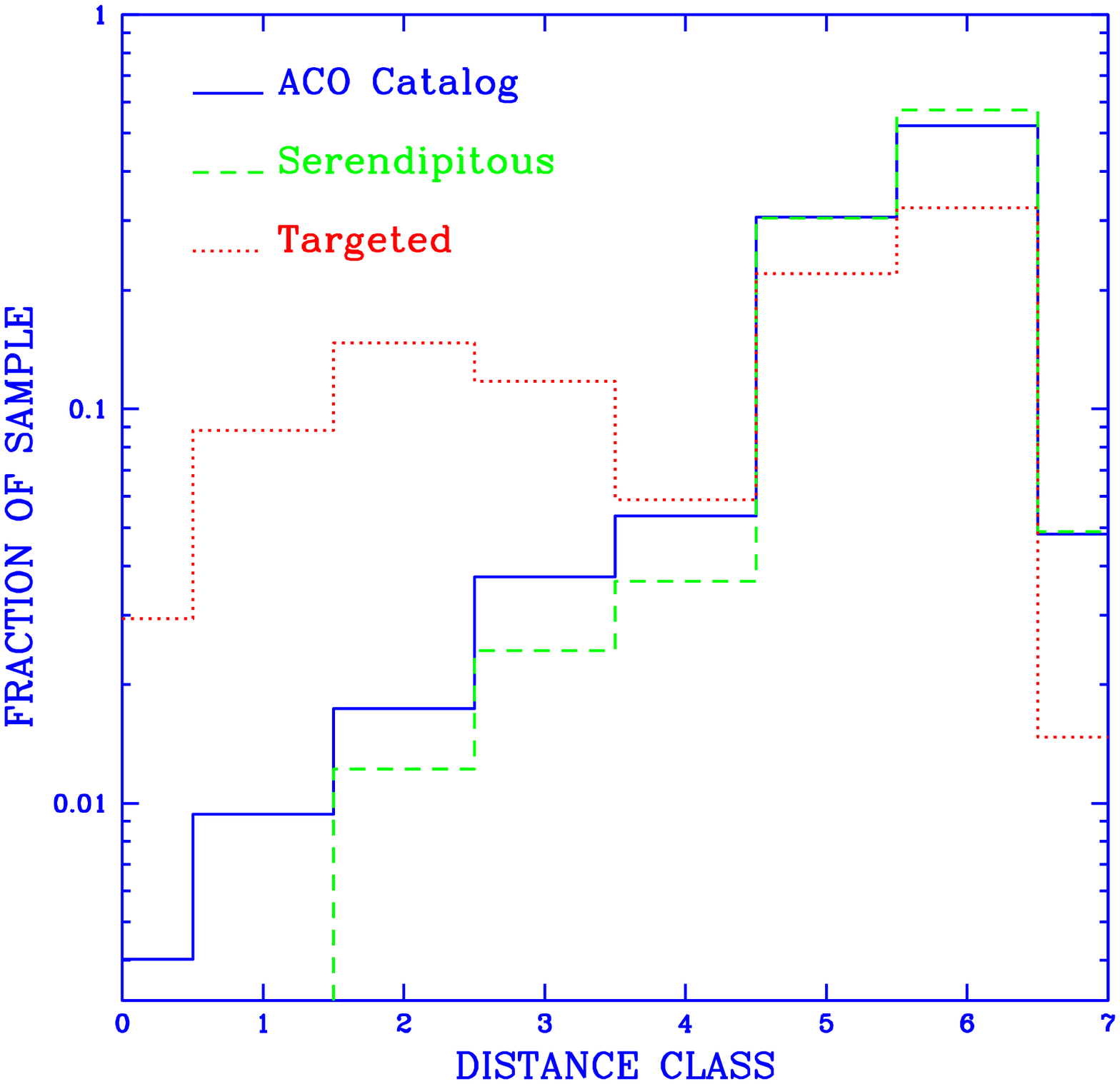}}


\rput[tl]{0}(0,8.0){
\begin{minipage}{18cm}
\small\parindent=3.5mm
{\sc Fig.}~1. Comparison of the optical properties of clusters in
the ACO catalog with the optical properties of clusters observed
with the PSPC. The PSPC sample is divided into serendipitously
observed clusters and targeted observations.
Comparisons are shown between a) cluster richness ($N_{gal}$),
b) Bautz Morgan Type, c) Rood Sastry Type, and d) distance class.
\end{minipage}
}
\endpspicture
\end{figure*}

\begin{figure*}[tb]
\pspicture(0,9.8)(18.5,22.9)

\rput[tl]{0}(4.0,27.2){\epsfxsize=8.5cm
\epsffile{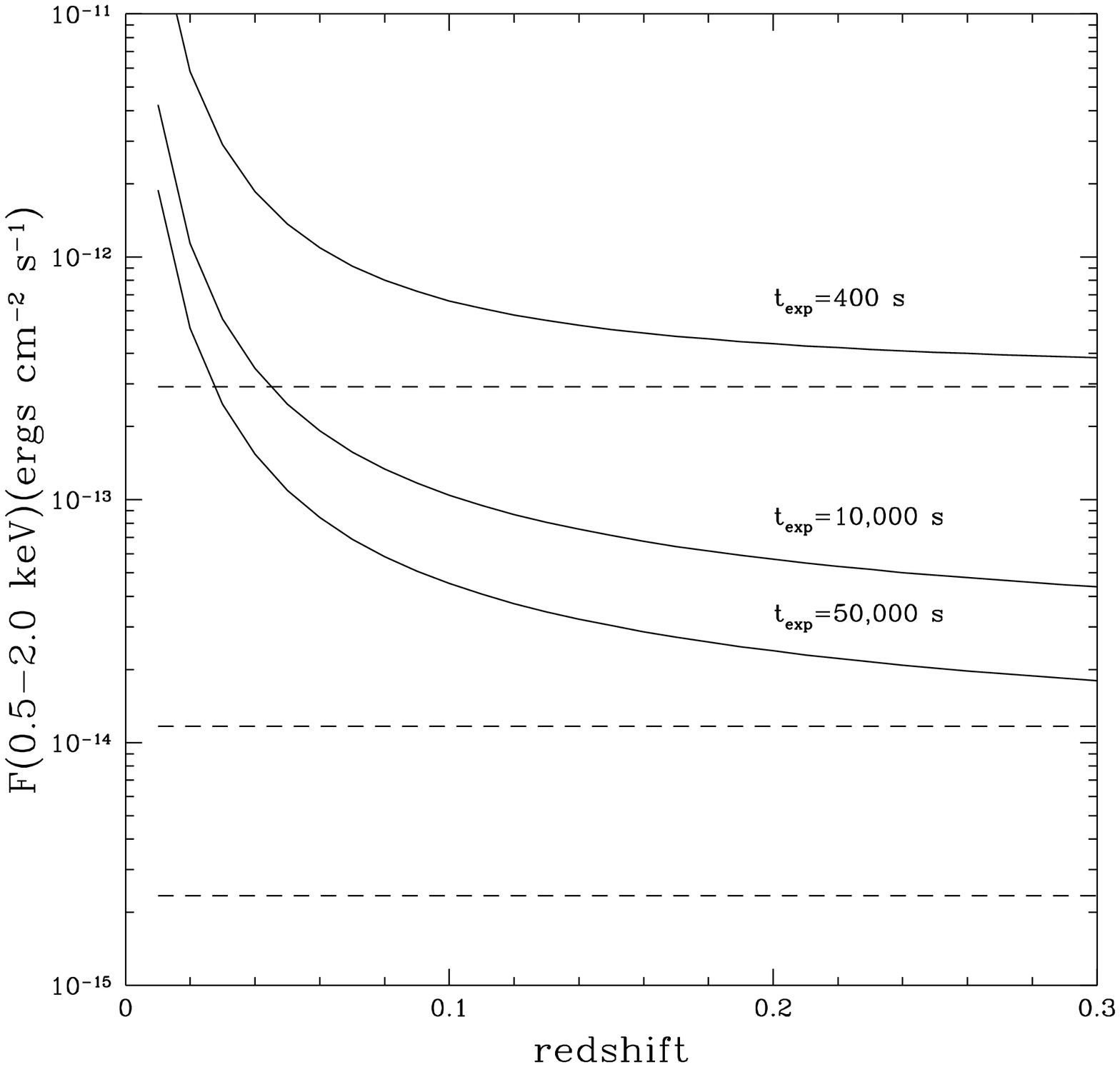}}

\rput[tl]{0}(0,18.0){
\begin{minipage}{18cm}
\small\parindent=3.5mm
{\sc Fig.}~2. Typical $3 \sigma$ flux limits for extended sources
(solid lines) at several representative PSPC exposure times as
a function of redshift
(see the text for details).  Also shown in the corresponding
$3 \sigma$ flux limit for point sources (dashed lines).
\end{minipage}
}

\rput[tl]{0}(4.00,16.2){\epsfxsize=7.75cm
\epsffile[30 470 530 678]{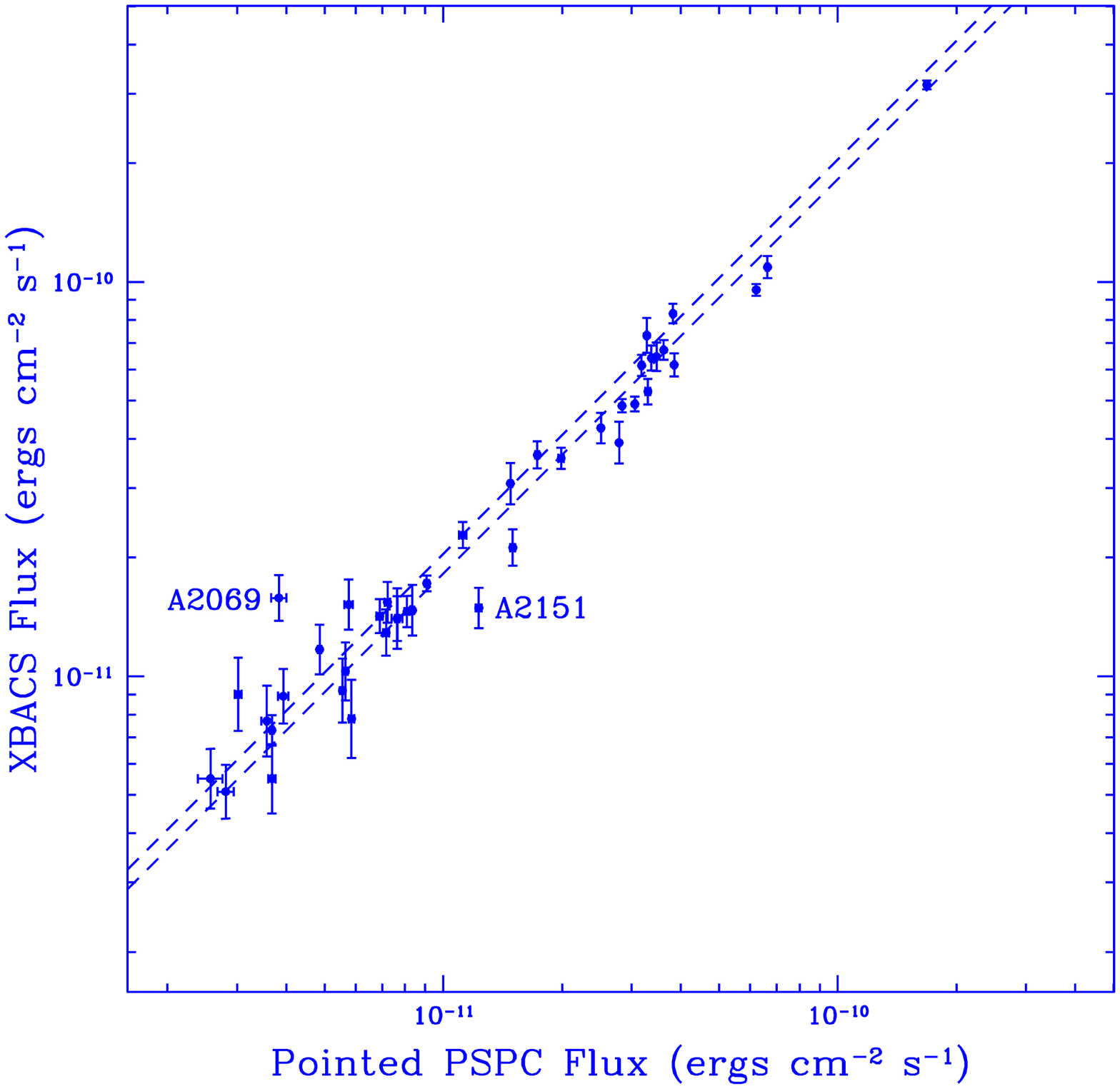}}

\rput[tl]{0}(0,7.0){
\begin{minipage}{18cm}
\small\parindent=3.5mm
{\sc Fig.}~3. A comparison of the measured fluxes for $R \geq 2$
clusters in XBACS that were also observed during pointed PSPC observations.
The XBACS flux is the total 0.1-2.4~keV flux given in Ebeling $\etal$
(1996) and the pointed PSPC flux is the 0.5-2.0~keV flux within the
central 1~Mpc listed in Table 3.  The dashed lines indicate
exact agreement under the assumption of a spherically symmetric
$\beta$ model for clusters with core radii of 100~kpc (lower line)
and 250~kpc (upper line). The two most discrepant clusters are
identified in the plot.
\end{minipage}
}

\endpspicture
\end{figure*}


\begin{figure*}[tb]
\pspicture(0,9.8)(18.5,22.9)

\rput[tl]{0}(4.0,27.2){\epsfxsize=8.5cm
\epsffile{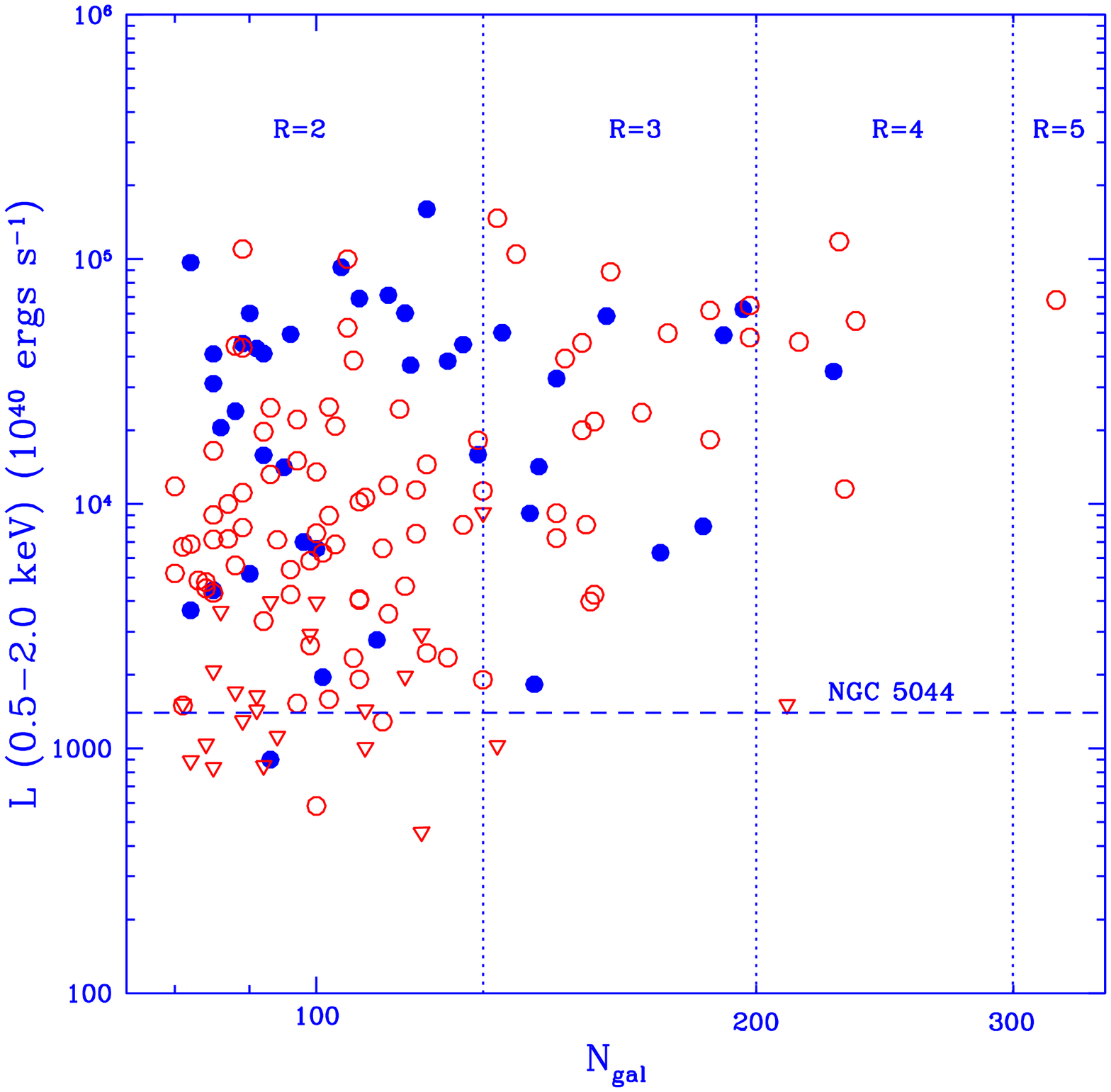}}

\rput[tl]{0}(0,18.0){
\begin{minipage}{18cm}
\small\parindent=3.5mm
{\sc Fig.}~4. Scatter plot of X-ray luminosity vs. cluster richness 
($N_{gal}$) for
the entire sample.  Early Bautz Morgan Type clusters are shown
as filled symbols and late Bautz Morgan Type clusters are shown
as open symbols. Detections are shown as circles and upper limits are
shown as triangles.
\end{minipage}
}

\rput[tl]{0}(4.00,16.2){\epsfxsize=7.75cm
\epsffile[30 470 530 678]{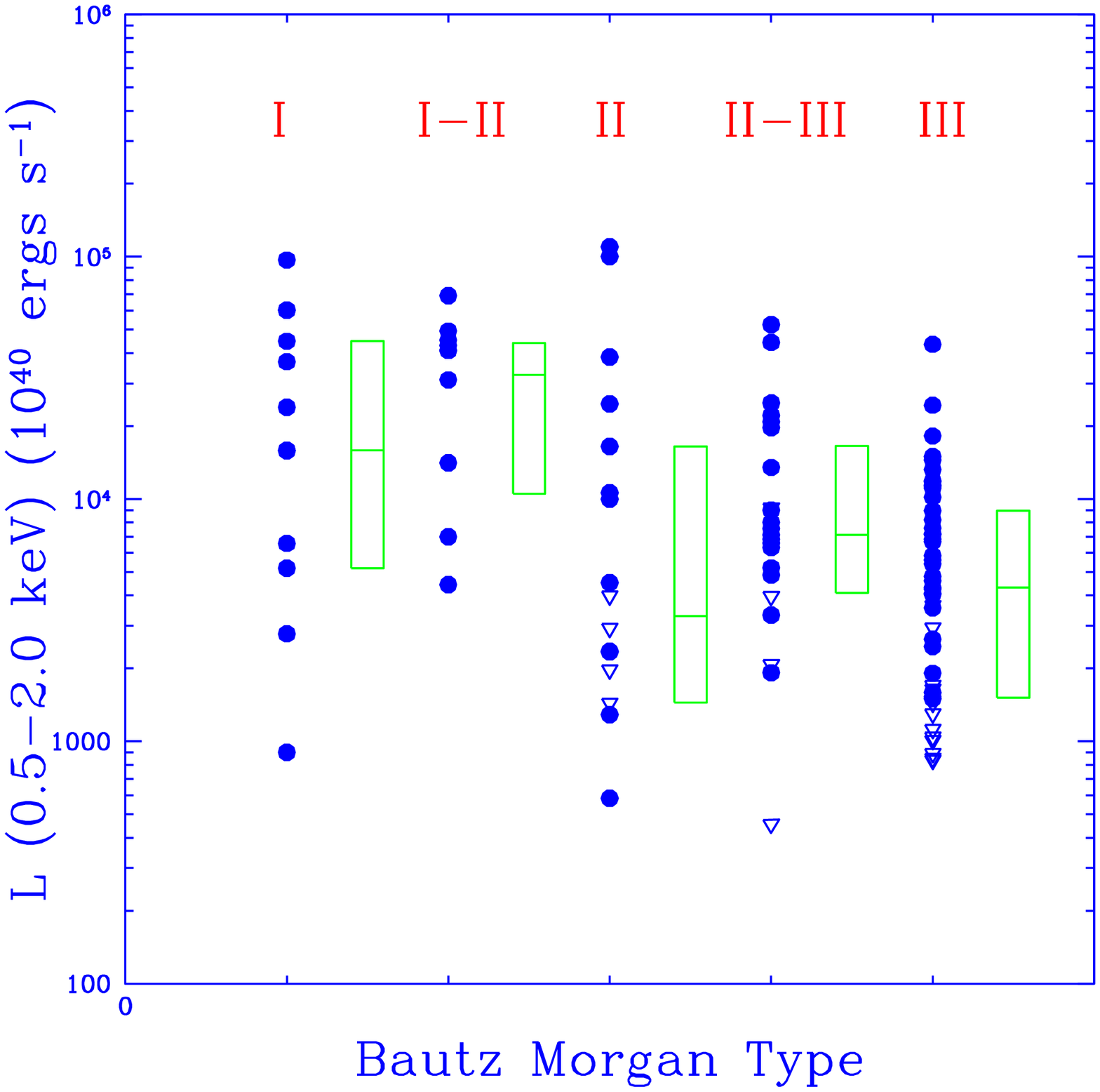}}

\rput[tl]{0}(0,7.0){
\begin{minipage}{18cm}
\small\parindent=3.5mm
{\sc Fig.}~5. Scatter plot of X-ray luminosity vs. Bautz Morgan Type
for R=2 clusters.
Detections are shown as circles and upper limits are
shown as triangles. Also shown are the 25, 50, and 75\% percentiles
for each Bautz Morgan Type.
\end{minipage}
}

\endpspicture
\end{figure*}

\begin{figure*}[tb]
\pspicture(0,9.8)(18.5,22.9)

\rput[tl]{0}(0.5,27.2){\epsfxsize=8.5cm
\epsffile{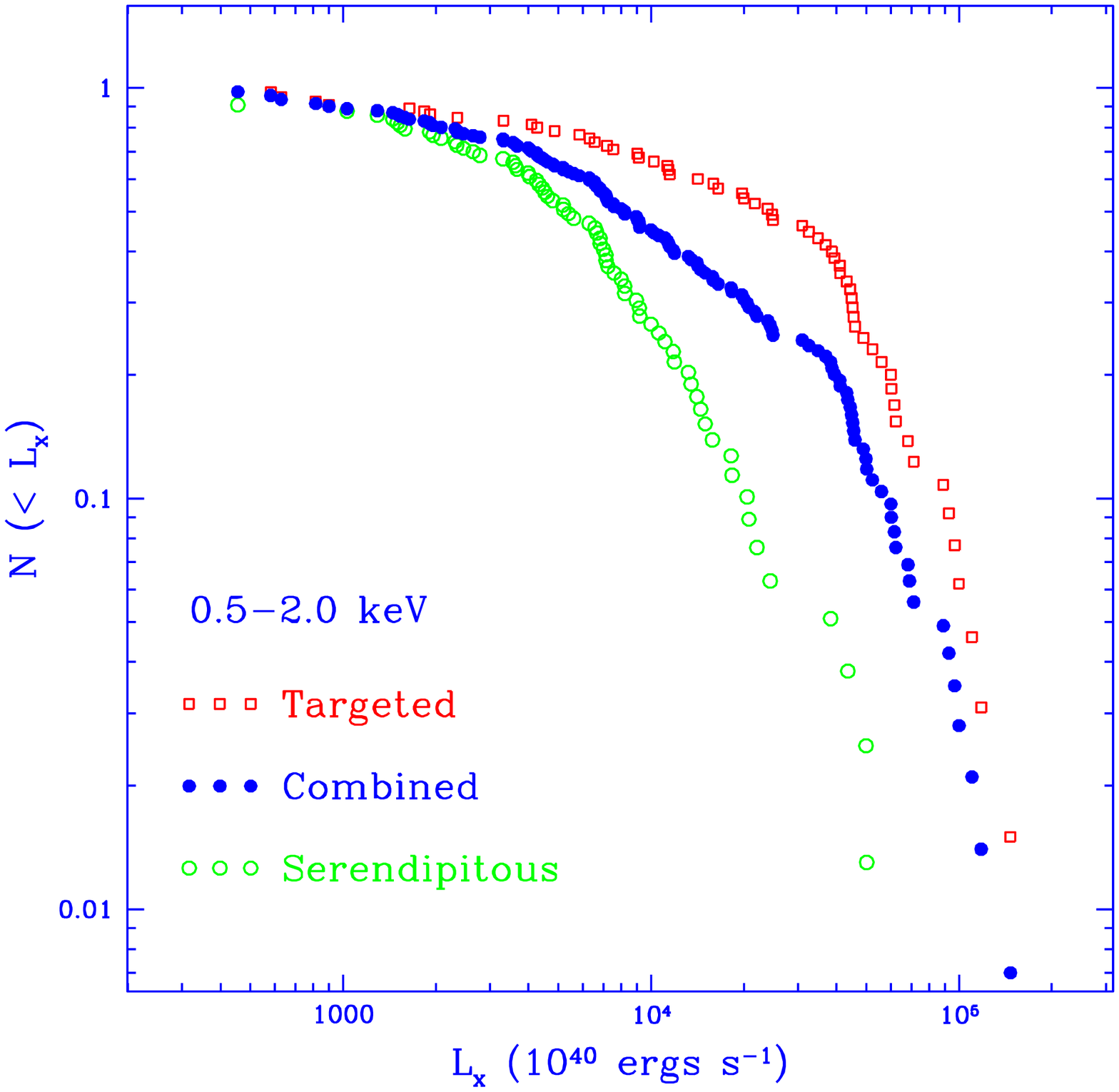}}

\rput[tl]{0}(0.75,16.2){\epsfxsize=7.75cm
\epsffile[30 470 530 678]{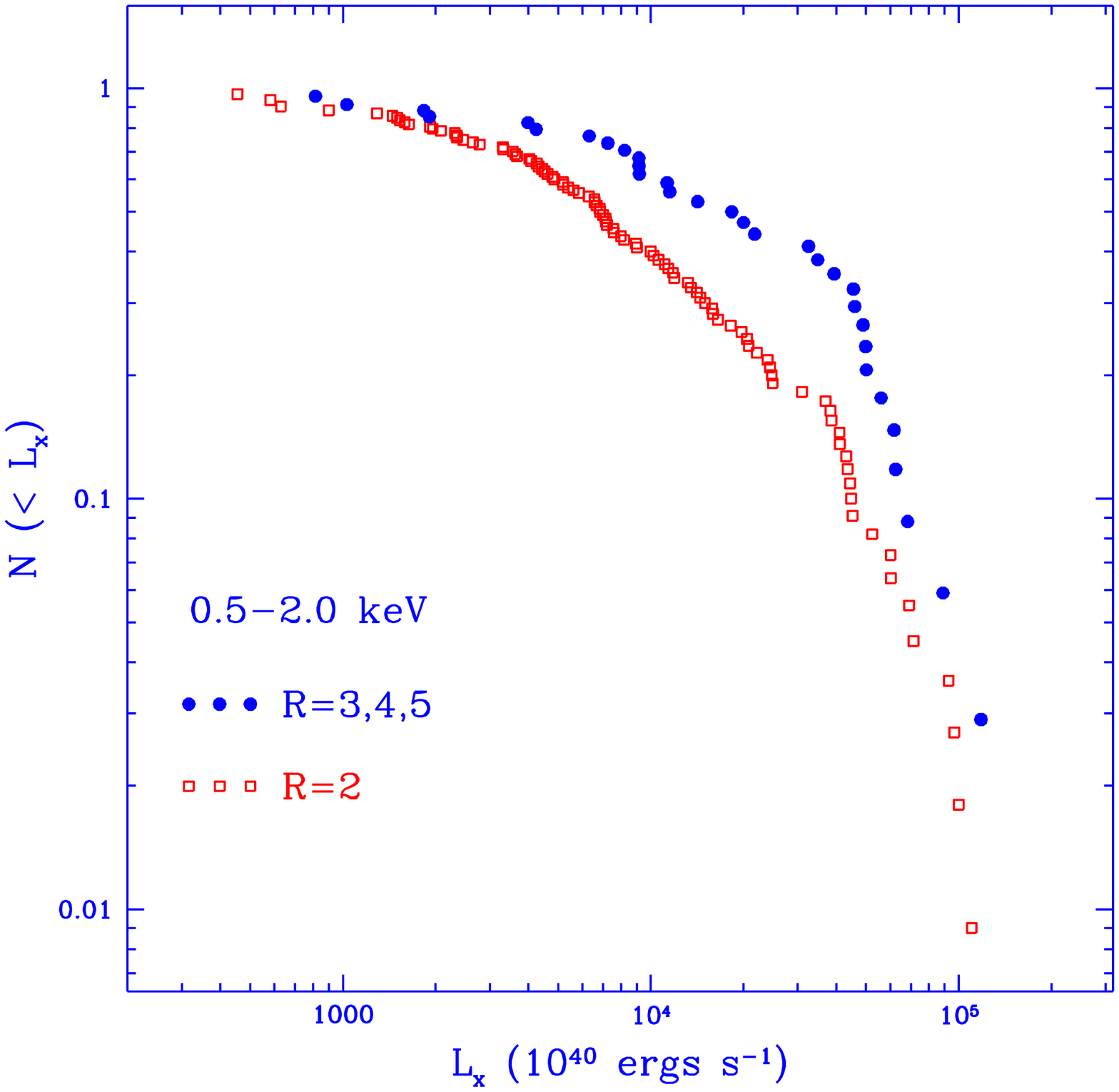}}

\rput[tl]{0}(9.5,28.2){\epsfxsize=8.5cm
\epsffile{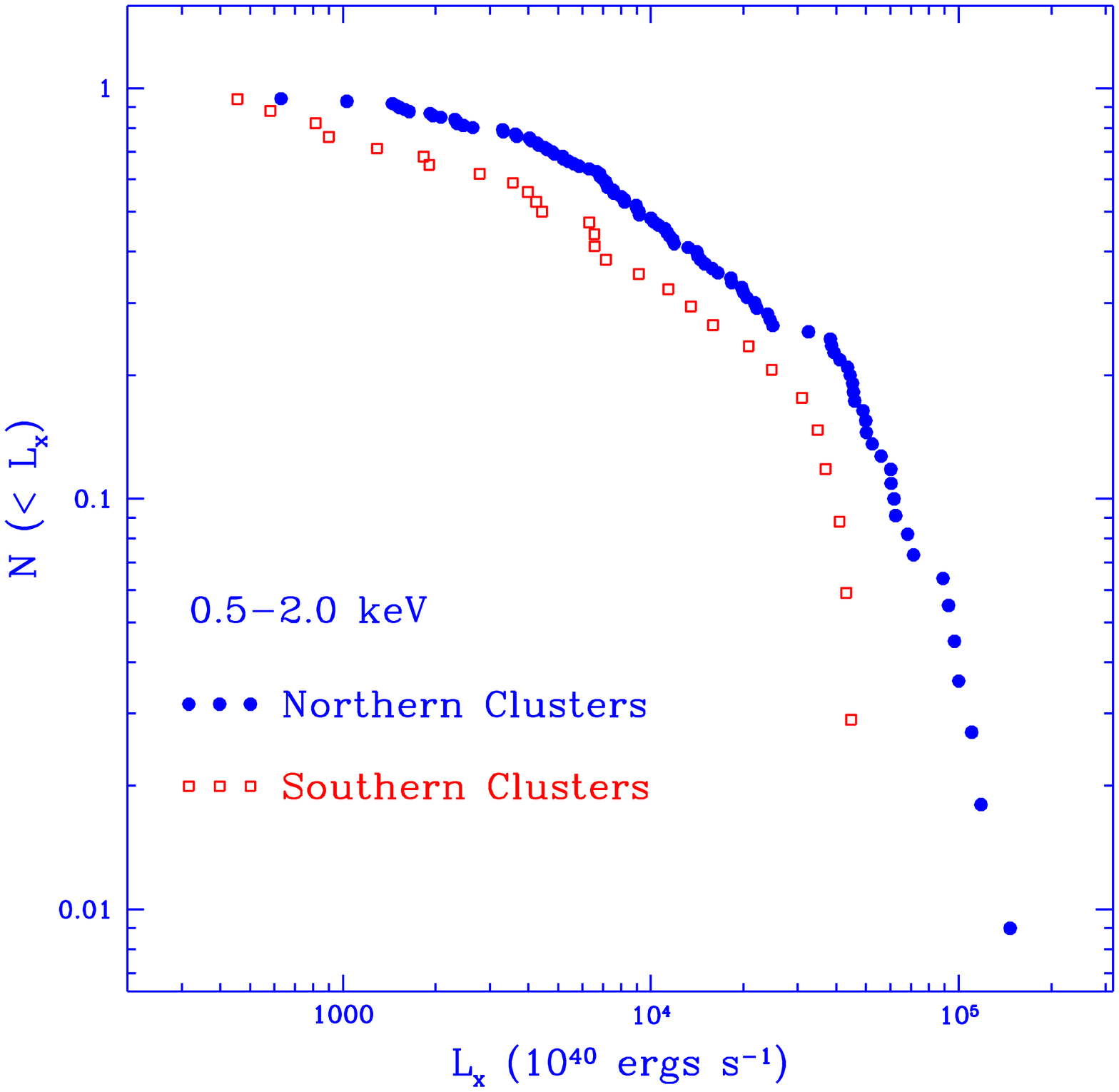}}

\rput[tl]{0}(9.75,16.2){\epsfxsize=7.75cm
\epsffile[30 470 530 678]{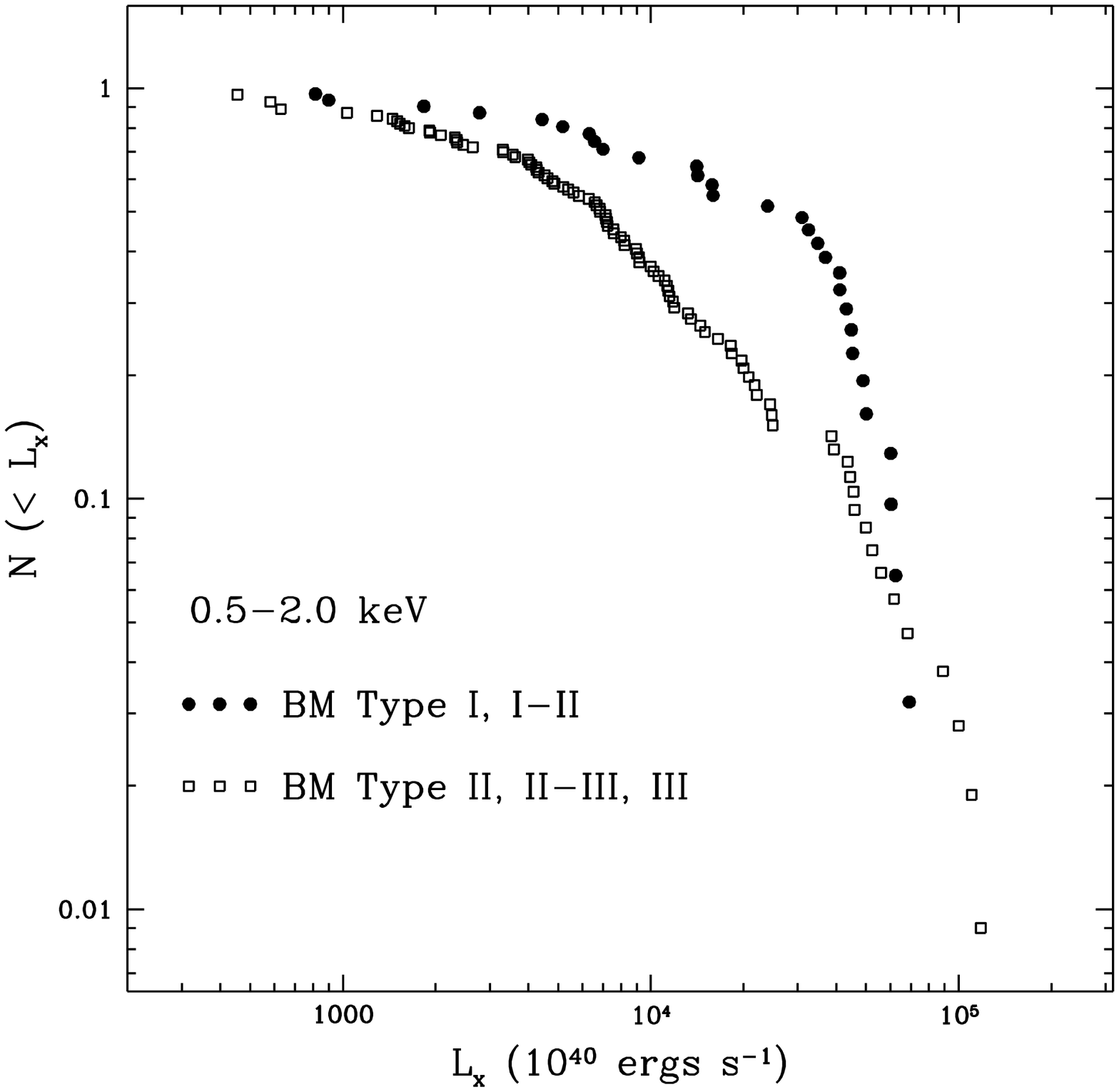}}


\rput[tl]{0}(0,7.0){
\begin{minipage}{18cm}
\small\parindent=3.5mm
{\sc Fig.}~6. Comparison of the 0.5-2.0~keV luminosity function for different
sub samples. a)  serendipitous vs. targeted samples, b) Abell vs.
southern ACO, c) richness class (R=2 vs. R=3,4, and 5), and
d) Bautz Morgan Types (early vs. late).
\end{minipage}
}
\endpspicture
\end{figure*}


\begin{figure*}[tb]
\pspicture(0,9.8)(18.5,22.9)

\rput[tl]{0}(0.5,27.2){\epsfxsize=8.5cm
\epsffile{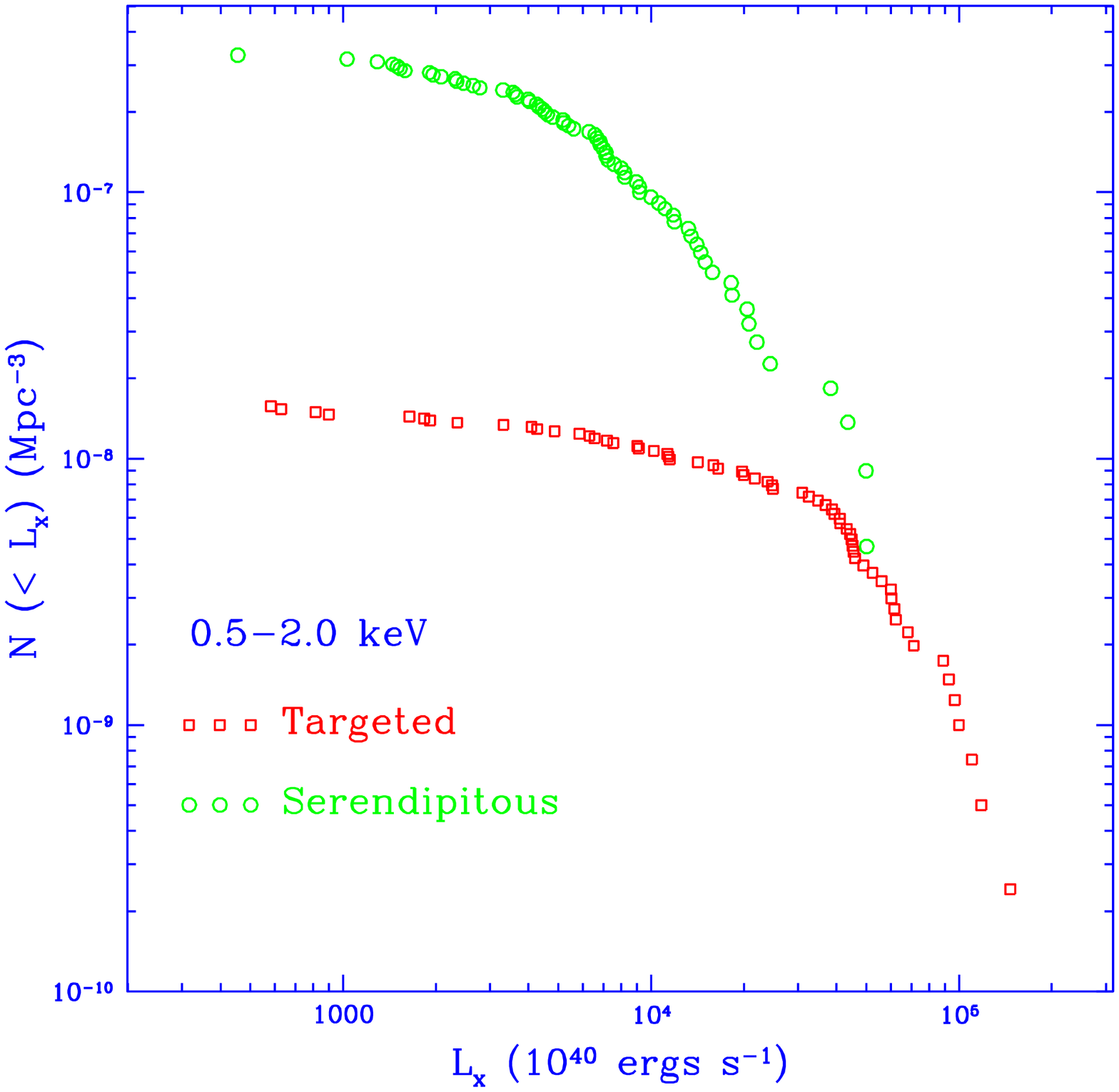}}

\rput[tl]{0}(9.5,27.2){\epsfxsize=8.5cm
\epsffile{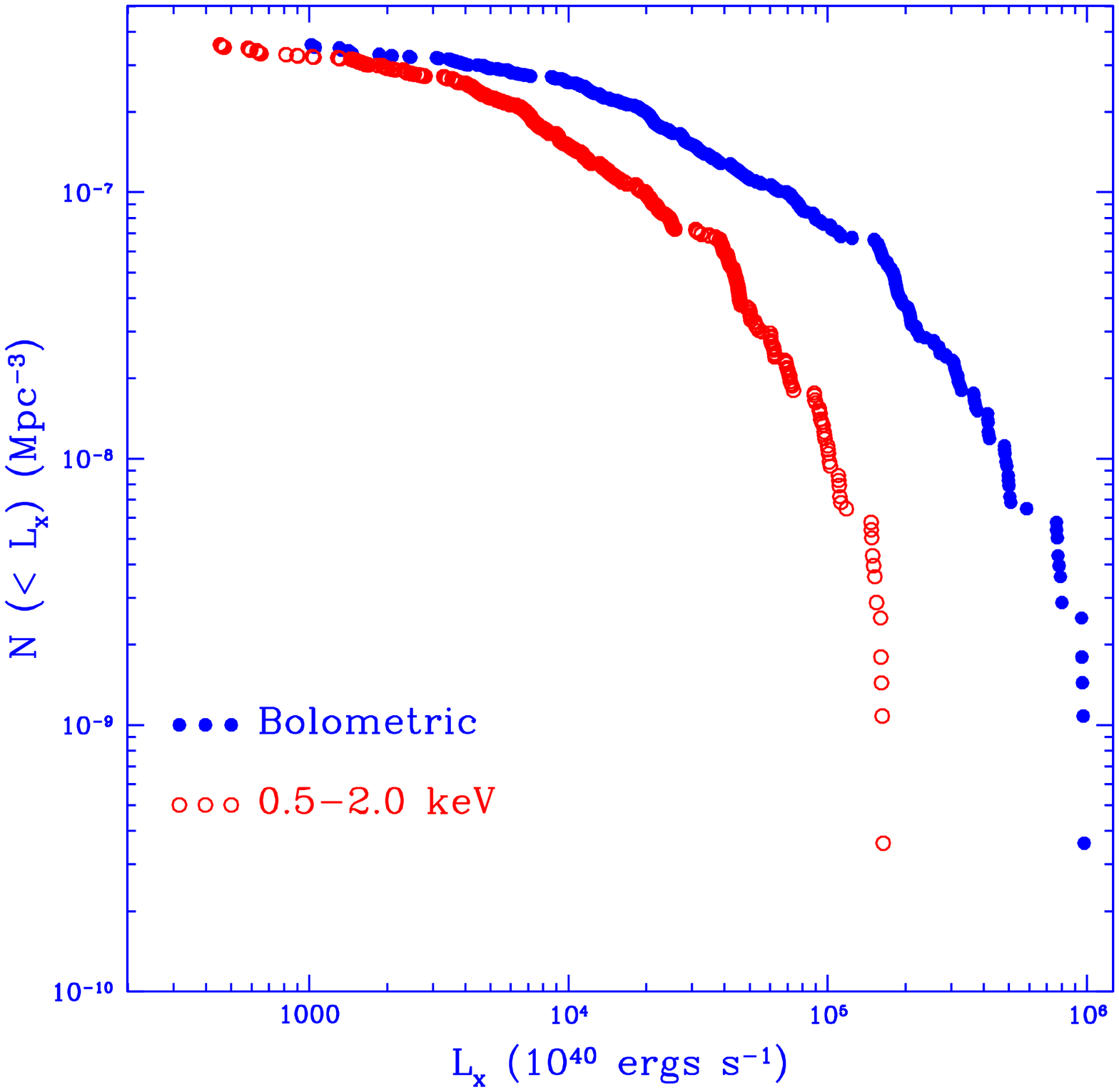}}

\rput[tl]{0}(0,18.0){
\begin{minipage}{8cm}
\small\parindent=3.5mm
{\sc Fig.}~7. Luminosity functions of the volume normalized
targeted sample and ACO number density normalized
serendipitous sample.
\end{minipage}
}

\rput[tl]{0}(10,18.0){
\begin{minipage}{8cm}
\small\parindent=3.5mm
{\sc Fig.}~8. Bolometric and 0.5-2.0~keV luminosity function of the entire
PSPC sample normalized using the technique described in the text.
\end{minipage}
}

\rput[tl]{0}(5.00,16.2){\epsfxsize=7.75cm
\epsffile[30 470 530 678]{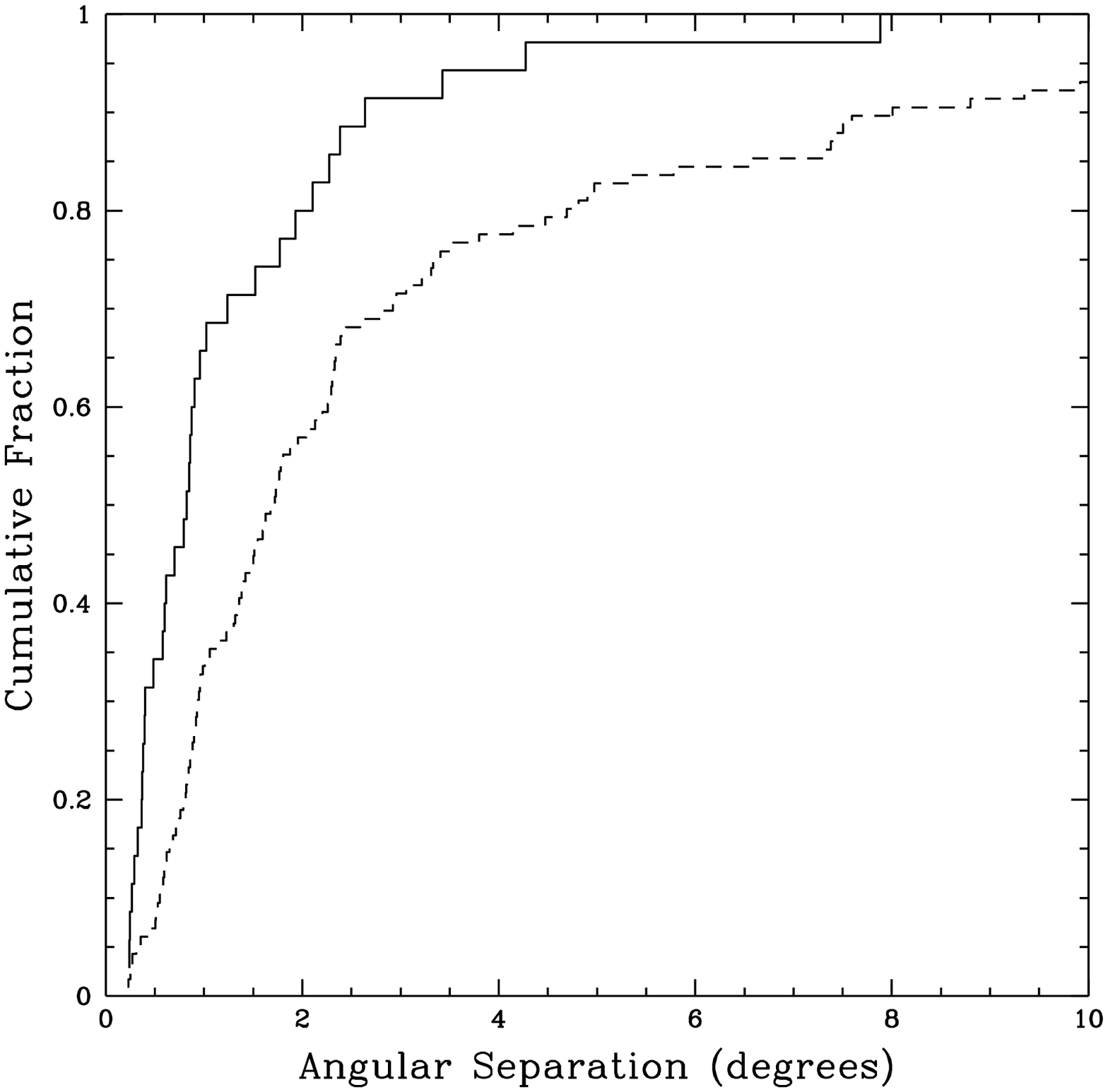}}

\rput[tl]{0}(0,7.0){
\begin{minipage}{18cm}
\small\parindent=3.5mm
{\sc Fig.}~9. The cumulative distribution of the angular separations
between clusters in the PSPC sample and their nearest neighboring
Abell or ACO cluster.  Only clusters with $\Delta m_{10} \leq 0.2$
are considered. The solid line
corresponds to low X-ray luminosity clusters
($L(0.5-2.0~keV) < 3.0 \times 10^{43} \ergs$) and the dashed lines
corresponds to more luminous clusters.
\end{minipage}
}
\endpspicture
\end{figure*}


\end{document}